\documentclass[preprint,12pt]{elsarticle}

\usepackage{amssymb,cprotect}
\newcounter{bla}

\journal{Computer Physics Communications}

\usepackage{cmap}
\usepackage[T1]{fontenc}
\usepackage[utf8]{inputenc}

\usepackage{amssymb,amsmath,bigfoot,color,graphicx}

\newcommand{\tmdice}{{\sf TMDICE}}
\newcommand{\mincas}{{\sf MINCAS}}
\newcommand{\jetmed}{{\sf JetMed}}

\usepackage[left=2cm, right=2cm, top=2cm, bottom=2cm]{geometry}

\usepackage{xcolor}

\begin{document}

\begin{frontmatter}

\title{The \tmdice\, Monte Carlo shower program and algorithm for jet-fragmentation via coherent medium induced radiations and scattering}


\author{Martin Rohrmoser\corref{author}}

\cortext[author] {Corresponding author.\\\textit{E-mail address:}  rohrmoser.martin1987@gmail.com}
\address{Cracow University of Technology, Faculty of Materials Engineering and Physics, ul. Podchorą\.{z}ych 1, 30-084 Krak\'ow, Poland,
Institute of Nuclear Physics, Polish Academy of Sciences,\\
  ul.\ Radzikowskiego 152, 31-342 Krak\'ow, Poland}

\begin{abstract}
Parton jets in the hot and dense medium of a Quark Gluon Plasma (QGP) can undergo multiple processes of scatterings off medium particles as well as processes of coherent medium induced radiations. A Monte-Carlo algorithm and resulting program is presented that allows to obtain jets that were formed by these two types of processes from an initial highly energetic quark or gluon. 
The program accounts for the increase in the momentum components of jet-particles transverse to the jet-axis due to processes of scattering as well as medium induced radiations in addition to energy-loss due to the medium induced radiations. 

\end{abstract}


\end{frontmatter}

% \linenumbers

% All CPiP articles must contain the following
% PROGRAM SUMMARY.

{\bf PROGRAM SUMMARY}
  %Delete as appropriate.

\begin{small}
\noindent
{\em Program Title: \tmdice}                                          \\
{\em CPC Library link to program files:} (to be added by Technical Editor) \\
{\em Developer's repository link:} (if available) \\
{\em Code Ocean capsule:} (to be added by Technical Editor)\\
{\em Licensing provisions(please choose one):} GPLv3  \\
{\em Programming language: C++, Bash}                                   \\
{\em Supplementary material:}                                 \\
  % Fill in if necessary, otherwise leave out.
{\em Journal reference of previous version:}*                  \\
  %Only required for a New Version summary, otherwise leave out.
{\em Does the new version supersede the previous version?:}*   \\
  %Only required for a New Version summary, otherwise leave out.
{\em Reasons for the new version:*}\\
  %Only required for a New Version summary, otherwise leave out.
{\em Summary of revisions:}*\\
  %Only required for a New Version summary, otherwise leave out.
{\em Nature of problem(approx. 50-250 words):}\\
In order to describe the fragmentation of parton cascades/jets in the medium processes of coherent medium induced radiation, where a particle emission is formed simultaneously to multiple scatterings off medium particles, need to be considered (in addition to scatterings off medium particles without emissions)[1,2,3,4,5,6,7].
A description of jet-fragmentation in the medium needs to be found that uses the effective kernels for coherent medium induced radiation and scattering [8,9] and which provides distributions of jet-particles  as a function of the time the jet needs for passing the medium.\\
  %Describe the nature of the problem here. \\
{\em Solution method(approx. 50-250 words):}\\
A Monte-Carlo method is presented that allows to obtain a set of jet particles from a predefined initial particle.
To this end, the variables relevant for the description of jet particles (such as the time of emission, momentum fraction and momentum component transverse to jet axis, and if the parton is a quark/antiquark or a gluon) are sampling from probability density functions that were obtained from the kernels for coherent medium induced radiation and scattering off medium particles in [8,9]. 
This is achieved in a two step process: First, before obtaining jet particles, the corresponding cumulative distribution functions are calculated and its inverse of the cumulative distribution functions obtained numerically. Then, samples are obtained by random selection from the inverse of the cumulative distribution functions.
  %Describe the method solution here.
{\em Additional comments including restrictions and unusual features (approx. 50-250 words):}\\
The kernels [8,9] for the coherent medium induced radiations were derived within the eikonal approximation that only momentum components transverse to the incoming particles are affected by medium transfers. Furthermore, these kernels do not depend on the time of emission (thus, neglecting effects of the finite size of the medium within the emission and scattering kernels).
For simplicity so far only a medium with constant parameters for the jet-medium interactions have been assumed. 
  %Provide any additional comments here.
   \\

* Items marked with an asterisk are only required for new versions
of programs previously published in the CPC Program Library.\\
\end{small}

\section{Introduction}
Monte-Carlo algorithms for the random generation of parton cascades/jets have been an important tool theoretical and phenomenological studies on jet production in ultrarelativistic nuclear collisions.
One reason is that they allow to obtain directly numerical results for the four-momenta of all of the particles of a jet, which was selected by a Monte-Carlo algorithm. 
This kind of information can then be used to obtain predictions for jet-observables -- by definition multiparticle observables -- such as jet-shapes, di-jet correlations, particle correlations between jet-particles, etc.

While for parton jets that fragment in vacuum (e.g. jets produced in proton-proton collisions) the evolution can be sufficiently described by Monte-Carlo algorithms that resum the collinear (by resummation via the Dokshitzer, Gribov, Lipatov, Altarelli, Parisi (DGLAP) evolution equations~\cite{ALTARELLI1977298,Dokshitzer:1977sg,Gribov:1972ri,Lipatov:1974qm}) and soft singularities of the bremsstrahlung processes, the situation for jet-fragmentation in the medium is less clear as several types of processes and effects may contribute: 
Main contributions are given by scatterings and medium induced radiations.
A large number of Monte-Carlo event generators~\cite{Armesto:2009fj, Schenke:2009gb,Zapp:2008gi,Zapp:2012ak,PhysRevC.78.034908,PhysRevC.88.014905,Lokhtin:2005px} exist that implement these processes of jet-fragmentation.
Processes of particle radiation off jet-particles in the medium can occur simultaneously with scatterings off medium particles, which give rise to interference effects that were first described in the context of QCD by Baier, Dokshitzer, Mueller, Peign\'e, Schiff, and independently by Zakharov (BDMPS-Z)~\cite{Baier:2000mf,Baier:2000sb,Zakharov:1996fv,Zakharov:1997uu,Zakharov:1999zk,Baier:1994bd,Baier:1996vi}.
BDMPS-Z found that the interferences between scatterings and medium induced radiation lead to a suppression of the emission of highly energetic jet-particles, an effect that was considered by several Monte-Carlo event generators, e.g.~\cite{Armesto:2009fj,Zapp:2008gi,Zapp:2012ak,Lokhtin:2005px}. A probability distribution in the form of an effective splitting kernel for medium induced coherent radiation that reproduces the BDMPS-Z emission spectra was found by Blaizot, Iancu, Dominguez, and Mehtar-Tani (BDIM)~\cite{Blaizot:2012fh} for highly energetic gluons in the medium and lead to the derivation of an integro-differential evolution equation~\cite{Blaizot:2013vha} for the in-medium evolution of gluon-jets undergoing medium induced coherent branchings as well as scatterings off medium particles.
A generalization of this kind of evolution equation to a set of evolution equations that describe the in-medium evolution of quarks in addition to gluons as jet particles has been formulated in~\cite{Mehtar-Tani:2018zba} for the simplified case of collinear parton emission without scatterings off medium particles and for the more general case that considers non-collinear parton emissions as well as scatterings in~\cite{Blanco:2021usa}.
A Monte-Carlo algorithm that describes jet-evolution via coherent medium induced splittings of quarks and gluons is the \jetmed\, algorithm~\cite{caucal:tel-03081993}. There, the branchings were considered as collinear and transverse momenta were selected from a Gaussian distribution.
A Monte-Carlo algorithm that provides a solution for the system of evolution equations given in~\cite{Blaizot:2013vha,Mehtar-Tani:2018zba,Blanco:2021usa} is the \mincas-program~\cite{Kutak:2018dim,Blanco:2020uzy,Blanco:2021usa}: 
It provides fragmentation functions for both the gluons as well as the quarks and considers non-collinear in-medium branchings and scatterings.

In this paper, I present a Monte-Carlo algorithm for transverse momentum dependent induced coherent emissions (\tmdice) that obtains momentum components of jet-particles using the splitting (for non-collinear coherent medium induced emissions) and scattering kernels from~\cite{Blanco:2021usa}.  
While there exists with \mincas\, a Monte-Carlo algorithm for the solution of the BDIM evolution equations, that algorithm does not provide momentum components (given via momentum fractions and momentum components transverse to the jet axis) for 
the individual jet particles (but rather samples of the fragmentation functions). 
\tmdice\, was created to fill this gap.
While other algorithms also study the production and evolution of jet-particles in the medium, to my knowledge \tmdice\, is also the first algorithm that uses the splitting and scattering kernels of~\cite{Blaizot:2012fh,Blaizot:2013vha,Blanco:2021usa} in order to describe in-medium jet-fragmentation.

This paper is organized as follows:
Sec.~\ref{sec2} describes the problem that the Monte-Carlo algorithm solves and the algorithm used for the solution.
Sec.~\ref{sec3} describes how to install and use the program.
Sec.~\ref{sec_ex} shows an example of results obtained from output of the data.
Sec.~\ref{sec4} concludes this paper.
\ref{appA} establishes the relation of the \tmdice\, algorithm to direct Monte-Carlo solutions of the BDIM equations of~\cite{Blaizot:2013vha,Blanco:2021usa}, such as, e.g. \mincas.

\section{Formalism}
\label{sec2}
The \tmdice\, Monte-Carlo algorithm describes the fragmentation of a highly energetic initial particle into a jet via processes of both coherent medium induced splittings and scatterings off medium particles that correspond to the ones in~\cite{Blaizot:2012fh,Blaizot:2013vha,Blanco:2021usa}. 
The splitting kernels presented therein where obtained for a time-independent medium. 
Therefore, the here presented \tmdice-Monte-Carlo algorithm will also assume a time-independent medium. 
This section describes how the \tmdice-algorithm for in-medium jet-fragmentation is inferred from the splitting and scattering kernels. The appendix~\ref{appA} shows that the the resulting multiplicity densities of jet particles solve a integro-differential evolution equation and the corresponding fragmentation functions solve the integro-differential evolution equation deduced in~\cite{Blaizot:2013vha} and solved directly by the \mincas-algorithm~\cite{Kutak:2018dim,Blanco:2020uzy,Blanco:2021usa}.

The probability density $\mathcal{P}_{BA}$ for a coherent medium induced parton branching of a particle $A$ into two particles $B$ and $C$ at time $t$, where $k_{A+}$, $k_{B+}$, and $k_{C+}$ are the light cone energies and $\mathbf{k}_{A}$, $\mathbf{k}_{B}$, and $\mathbf{k}_{C}$ are the transversal components of the particle momenta, is given with~\cite{Blaizot:2012fh,Blaizot:2013vha,Blanco:2021usa}
\begin{equation}
    \frac{\partial^4 \mathcal{P}_{BA}}{\partial t \partial z \partial^2 \mathbf{Q} }=\frac{\alpha_s}{(2\pi)^2}\mathcal{K}_{BA} (\mathbf{Q},z, k_{A+})\,,
\end{equation}

via the splitting kernels for coherent medium induced radiation 
\begin{equation}
{\cal K}_{BA}(\mathbf{Q},z,k_{A+})=\frac{2}{k_{A+}}\frac{P_{BA}(z)}{z(1-z)}\sin\left[\frac{Q^2}{2k_{br}^2}\right]\exp\left[-\frac{Q^2}{2k_{br}^2}\right] \,,
\label{eq:Kqz}
\end{equation}
\begin{equation}
k_{br}^2=\sqrt{\omega_0\hat q_0},\,\,\,\,\,\,\,\, \omega_0=z(1-z)k_{A+}
\end{equation}
\begin{equation}
\hat q_0=\frac{\hat q}{N_c} f_{BA}(z)\,,  
\end{equation}
where $N_c$ is the number of colors. $\hat q$ is considered as
\begin{equation}
\hat{q}=\frac{\partial \langle k_\perp^2 \rangle}{dt}\,,
\end{equation} 
where $k_\perp$ is the momentum component in direction transverse to the incident jet particle that is transferred to the jet particle by the medium. In the current version of the algorithm $\hat{q}$ was assumed as a time-independent constant.
The variables $z$ and $\mathbf{Q}$ are defined as follows:
\begin{align}
    z&=\frac{k_{B+}}{k_{A+}}\,,&1-z&=\frac{k_{C+}}{k_{A+}}\,,\\
    \mathbf{k}_{B}&=z\mathbf{k}_{A}+\mathbf{Q}\,,&\mathbf{k}_{C}&=(1-z)\mathbf{k}_{A}-\mathbf{Q}\,.
\end{align}
$P_{BA}(z)$ are the leading order DGLAP splitting functions for a parton $A$ that produces a parton $B$.
The functions $f_{BA}(z)$ are 
\begin{align}
    f_{gg}(z) &= (1-z) C_A + z^2 C_A \,, \\
    f_{qg}(z) &= C_F - z(1-z) C_A \,,\\
    f_{gq}(z) &= (1-z) C_A + z^2 C_F \,,\\
    f_{qq}(z) &= z C_A + (1-z)^2 C_F \,.
\end{align}
For medium induced scattering at time $t$ the probability density $\mathcal{P}_{A}$ is given via a scattering kernel $w_A(\mathbf{q})$, with $q$ the transverse momentum transferred to the jet particle, as
\begin{equation}
    \frac{\partial^3 \mathcal{P}_{A}}{\partial t \partial^2 \mathbf{q}}=\frac{1}{(2\pi)^2}w_A(\mathbf{q})\,,
    \label{eq:scatkerneldef}
\end{equation}
where the scattering kernels of quarks $w_q$ are related to the ones of gluons, $w_g$ as
\begin{equation}
w_q(\mathbf{q})=\frac{C_F}{C_A}w_g(\mathbf{q})\,,
\label{eq:relation_q_to_g}
\end{equation}
and the function $w_g(\mathbf{q})$ 
was considered as either~\cite{Blaizot:2014rla}:
\begin{equation}
 w_g(\mathbf{q}) = \frac{16\pi^2\alpha_s^2N_cn_{\rm med}}{\mathbf{q}^4}\,,
\label{eq:wq1}
\end{equation}
where $n_{\rm med}$ is the density of scatterers in the medium,
or~\cite{Gyulassy:1993hr}
\begin{equation}
 w_g(\mathbf{q}) = \frac{g^2m_D^2T}{\mathbf{q}^2(\mathbf{q}^2+m_D^2)}\,,
 \qquad g^2 = 4\pi\alpha_s\,,
\label{eq:wq3}
\end{equation}
with the Debye mass $m_D$ of the medium,
which alternatively can also be given in the following form:
\begin{equation}
 w_g(\mathbf{q}) = \frac{16\pi^2\alpha_s^2N_cn_{\rm med}}{\mathbf{q}^2(\mathbf{q}^2+m_D^2)}\,.
\label{eq:wq2}
\end{equation}
In the splittings and scatterings of Eqs.~(\ref{eq:Kqz}), (\ref{eq:wq1}), (\ref{eq:wq3}), and (\ref{eq:wq2}) the running of the coupling $\alpha_s$ with the involved momentum scales was not considered. Thus,$\alpha_s$ is up to now considered as a constant in the \tmdice~algorithm. Furthermore, the parameters that describe the medium in Eqs.~(\ref{eq:Kqz}), (\ref{eq:wq1}), (\ref{eq:wq3}), and (\ref{eq:wq2}), $\hat{q}$, $n_{\rm med}$, $m_D$, and $T$ were considered as  constants, due to the assumption of a time-independent medium.

The probability densities for coherent medium induced radiation and scattering can be resummed to yield the probility density that a jet-particle $A$  
which exists at time $t_1$ does not undergo any jet-medium interactions until a time $t_2$. The result is the following Sudakov-factor
\begin{equation}
    \Delta_A(x_A,t_2-t_1)=\exp{\left(-\Phi_A(x_A)(t_2-t_1)\right)}\,,
    \label{eq:sud}
\end{equation}
where the dependence on the light-cone energy of the considered jet particle $k_{\rm A+}$ is given via a light-cone energy fraction $x_A$ defined as 
\begin{equation}
x_A=\frac{k_{\rm A+}}{p_+}\,,
\end{equation}
where $p_+$ is the light-cone energy of the initial jet particle, from which all other jet-particles are obtained via processes of scattering and splitting. The functions $\Phi_g(x_A)$ and $\Phi_q(x_A)$ are 
\begin{align}
    \Phi_g(x_A)=&\alpha_s\int_\epsilon^{1-\epsilon} dz\int_{q>0} \frac{d^2\mathbf{q}}{(2\pi)^2}\bigg[\mathcal{K}_{gg}(\mathbf{q},z,x_Ap_+)+\mathcal{K}_{qg}(\mathbf{q},z,x_Ap_+)\bigg]+\int_{q>q_{\rm min}} \frac{d^2\mathbf{q}}{(2\pi)^2}w_g(\mathbf{q})\,,
    \label{eq:phig}\\
    \Phi_q(x_A)=&\alpha_s\int_\epsilon^{1-\epsilon} dz\int_{q>0}\frac{d^2\mathbf{q}}{(2\pi)^2}\mathcal{K}_{qq}(\mathbf{q},z,x_Ap_+)+\int_{q>q_{\rm min}} \frac{d^2\mathbf{q}}{(2\pi)^2}w_q(\mathbf{q})\,,
    \label{eq:phiq}
\end{align}
where the shorthand notation $\int_{q>b}d^2\mathbf{q}=\int_{q>b}^\infty q dq \int_{0}^{2\pi}d\phi_q$ (with $\phi_q$ the polar angle of $\mathbf{q}$) was used and $\epsilon$ and $q_{\rm min}$ are infrared regulators for the integrations over $z$ and $\mathbf{q}$ respectively.

Consequently, the probability density for a particle $A$, with momentum fraction $x_A$, present at time $t_1$ to split into particles $B$ and $C$ at time $t_2$, with momentum fraction $z$ and transferred transverse momentum $\mathbf{Q}$ is given by
\begin{equation}
    \Delta_A(x_A,t_2-t_1)\left( \frac{\alpha_s}{(2\pi)^2}\mathcal{K}_{BA}(\mathbf{Q},z,x_Ap_+)\right)\,,
    \label{eq:splitprob}
\end{equation}
and, analogously the probability density for a scattering of particle $A$ with transverse momentum transfer $\mathbf{q}$ at time $t_2$ from the medium is given by 
\begin{equation}
    \Delta_A(x_A,t_2-t_1)\left( \frac{1}{(2\pi)^2}w_A(\mathbf{q})\right)\,.
    \label{eq:scatprob}
\end{equation}
The main goal of the \tmdice-algorithm is to select for any jet-particle the possible interactions with the medium, according to Eqs.~(\ref{eq:splitprob}) and~(\ref{eq:scatprob}) and repeat this procedure for the particles resulting from the jet-medium interactions, until a jet formed between an initial time-scale $t_0$ and a final time-scale $t_{\rm max}$ has been found.
To this end, the above probability densities, Eqs.~(\ref{eq:splitprob}) and~(\ref{eq:scatprob}) are reformulated as
\begin{align}
   &
    \left(\Delta_A(x_A,t_2-t_1)
    \phi_A(x_A)\right)
    \left(
    \frac{\sum_B \rho_{BA}(x_A)}{\phi_A(x_A)}
    \right)
    \left(
    \frac{\rho_{BA}(x_A)}{\sum_B \rho_{BA}(x_A)}
    \right)
    \left(
    \frac{1}{\rho_{BA}(x_A)}
     \frac{\alpha_s}{(2\pi)^2}\mathcal{K}_{BA}(\mathbf{Q},z,x_Ap_+)\right)\,,
     \label{eq:splitprob2}
     \\
       &
    \left(\Delta_A(x_A,t_2-t_1)
    \phi_A(x_A)\right)
    \left(
    \frac{W_A}{\phi_A(x_A)}
    \right)
    \left(
    \frac{1}{W_A}
     \frac{1}{(2\pi)^2}w_A(\mathbf{q})\right)\,,
     \label{eq:scatprob2}
\end{align}
with
\begin{align}
    \rho_{BA}(x_A)=&\alpha_s\int_\epsilon^{1-\epsilon} dz\int_{q>0}\frac{d^2q}{(2\pi)^2}\mathcal{K}_{BA}(\mathbf{q},z,x_Ap_+)\,,\\
    W_A=&\int\frac{d^2\mathbf{q}}{(2\pi)^2}w_A(\mathbf{q})\,.
\end{align}
In Eqs.~(\ref{eq:splitprob2}) and~(\ref{eq:scatprob2}) the terms in the brackets can be identified as the following probabilities and probability distributions:
\begin{itemize}
    \item The probability distribution that a particle $A$ with momentum fraction $x_A$ given at time $t_1$ either splits or scatters at time $t_2$ is given by 
    \begin{equation}
        \phi_A(x_A)\Delta_A(x_A,t_2-t_1)\,.
        \label{eq:interactiontime}
    \end{equation}
    \item If it was found that a particle $A$ with momentum fraction $x_A$ undergoes a splitting or a scattering at time $t_2$, the conditional probability that a splitting occurs is given by $\frac{\sum_B\rho_{BA}(x_A)}{\phi_A(x_A)}$. Correspondingly, the conditional probability that a scattering occurs is given by 
    \begin{equation}
       1-\frac{\sum_B\rho_{BA}(x_A)}{\phi_A(x_A)}=\frac{W_A}{\phi_A(x_A)}\,.
       \label{eq:scat}
    \end{equation}
    \item 
    If a splitting happens to a particle $A$ with momentum fraction $x_A$ the type of splitting is determined 
    via the probability
    $\frac{\rho_{BA}(x_A)}{\sum_B\rho_{BA}(x_A)}$.
    Thus, if the splitting particle $A$ is a quark, the only possible type of splitting is a splitting into a quark and a gluon and, therefore it follows that 
    \begin{equation}
            \frac{\rho_{BA}(x_A)}{\sum_B\rho_{BA}(x_A)}=\frac{\rho_{qq}(x_A)}{\rho_{qq}(x_A)}=1\,.
            \label{eq:prob_qq}
    \end{equation}
    If the splitting particle $A$ is a gluon, then the possible types of splittings are either a splitting into a gluon pair or a quark-antiquark pair.
    Thus, with probability 
    \begin{equation}
            \frac{\rho_{BA}(x_A)}{\sum_B\rho_{BA}(x_A)}=\frac{\rho_{gg}(x_A)}{\rho_{gg}(x_A)+\rho_{qg}(x_A)}\,,
\label{eq:prob_gg}
    \end{equation}
    a splitting into a gluon pair occurs, while with probability 
    \begin{equation}
          \frac{\rho_{BA}(x_A)}{\sum_B\rho_{BA}(x_A)}=\frac{\rho_{qg}(x_A)}{\rho_{gg}(x_A)+\rho_{qg}(x_A)}=1-\frac{\rho_{gg}(x_A)}{\rho_{gg}(x_A)+\rho_{qg}(x_A)}\,,
          \label{eq:prob_gq}
    \end{equation}
 a splitting into a quark-antiquark pair occurs.
    \item If a particle $A$ with momentum fraction $x_A$ splits into particles $B$ and $C$, then the probability density for a splitting with momentum fraction $z$ and transverse momentum $\mathbf{Q}$ is given via 
    \begin{equation}
        \frac{\alpha_s  \frac{1}{(2\pi)^2}\mathcal{K}_{BA}(\mathbf{Q},z,x_Ap_+)}{\rho_{BA}(x_A)}\,.
        \label{eq:splitxq}
    \end{equation}
    \item If a scattering occurs for a particle of type $A$ the probability density for a scattering with transverse momentum transfer $\mathbf{q}$ is given by 
    \begin{equation}
        \frac{\frac{1}{(2\pi)^2} w_A(\mathbf{q})}{W_A}
        \label{eq:scatq}
    \end{equation}
\end{itemize}
Thus, with these probability distributions it is possible to formulate a Monte-Carlo algorithm that obtains all jet-particles with a certain minimal light cone energy (given by a minimal light cone energy fraction $x_{\rm min}$.
Here, we will only give the algorithm for the generation of a single jet from one initial jet-particle.
In order to obtain a statistically precise description of jet-observables, this Monte-Carlo event generation needs to be repeated multiple times.
In order to formulate the algorithm for any jet-particle $A$ a set of variables $v_A$ that describes this jet-particle shall be shortly defined:
For any particle $A$, the light cone energy, given by the corresponding energy fraction $x_A$, the transverse momentum components $\mathbf{k}_A$, and the time $t_A$ at which a jet-medium interaction occurs need to be known as well as whether the particle is a quark or a gluon, for which a variable $typ_A$ was defined. 
Thus, the following minimal set of variables $(v_A)_{\rm min}$ is necessary to suitably describe a particle $A$
\begin{equation}
    (v_A)_{\rm min}=\{typ_A, x_A, \mathbf{k}_A, t_A\}\,.
\end{equation}
In addition for practical convenience the \tmdice-algorithm also stores the time $t_{A\,{\rm old}}$ at which the particle was produced and, furthermore, a boolean variable $dump_A$ was defined, which is set to \verb#true# if the momentum fraction is below the infrared cutoff $x_A\leq x_{\rm min}$\footnote{While the algorithm does not consider particles with momentum fractions below $x_{\rm min}$ to participate in the further evolution of the jet -- by undergoing further scatterings or branchings -- it might be convenient in some cases to know the contributions from these very soft emissions, which is why these particles are stored with this extra flag.} and set to \verb#false# otherwise. 
Thus, in the \tmdice-algorithm, the following set of variables is used
\begin{equation}
    v_A=\{typ_A, x_A, \mathbf{k}_A, t_A,t_{A\,{\rm old}},dump_A\}\,.
\end{equation}

The principle of the Monte-Carlo algorithm is to obtain for an initial jet particle with variables $v_1$ that exists at time $t_1$ all the jet-particles $i\in \mathbb{N}$, represented via  observables $v_i$(with light-cone energy fractions $x_i\geq x_{\rm min}$ ), that were created from it as a result of scatterings and/or medium induced coherent emissions before a time $t_i\leq t_{\rm max}$. 
In order to specify the algorithm, one can define the sets
\begin{align}
    \mathcal{S}&=\{v_{1},\,v_{2},\,\dots v_{n}\}\,,&\\
    \mathcal{S}'&=\{v'_{1},\,v'_{2},\,\dots v'_{n'}\}\,,&\\
    \mathcal{S}_{\rm fin}&=\{v_{{\rm fin}\,1},\,v_{{\rm fin}\,2},\,\dots v_{{\rm fin}\,m}\}\,,&n,n',m\in\mathbb{N}\,.
\end{align}
The \tmdice-algorithm has the following general structure:
\begin{enumerate}
    \item An initial jet-particle with variables $v_1$ is determined.
    \item If $t_1\leq t_{\rm max}$ set $\mathcal{S}=\{v_1\}$ and $\mathcal{S}_{\rm fin}=\{\}$.
          Otherwise set $\mathcal{S}_{\rm fin}=\{v_1\}$ and $\mathcal{S}=\{\}$.
    \item If $\mathcal{S}\neq\{\}$ , a new set $\mathcal{S}'$ is created and $\mathcal{S}_{\rm fin}$ is modified (How these sets are created/modified is specified further below).
    \item If $\mathcal{S}'\neq\{\}$ set $\mathcal{S}=\mathcal{S}'$ and go to the previous step, otherwise the algorithm terminates.
\end{enumerate}
Step 1 in the above algorithm was treated as follows:
The variables $x_1$, $\mathbf{k}_1$, $typ_1$ as well as a timescale $t_0$ for the emission of this particle is set by the user. 
Then, the time $t_1$ is obtained by randomly selecting a number $R_{t_1}\in [0,1]$ from a uniform distribution and solving the equation 
    \begin{equation}
        R_{t_1}=\Delta_{typ_1}(x_1,t_{1}-t_0)\,.
        \label{eq:tsel}
    \end{equation}
Step 3 of the algorithm is carried out by performing the following steps for every particle $A$ (with variables $v_A$) in a non-empty set $\mathcal{S}$ (i.e. by executing a loop over the elements of $\mathcal{S}$):
\begin{enumerate}
    \item The type of interaction is determined. To this end, a number 
    $R_{int}\in [0,1]$ is randomly selected from a uniform distribution.
    If 
    \begin{equation}
           R_{int}<\frac{\sum_B\rho_{BA}(x_A)}{\phi_A(x_A)}
           \label{eq:spvssc}
    \end{equation}
    the particle undergoes a splitting, otherwise a scattering.
    \item  If jet-particle $A$ undergoes a scattering -- with outgoing jet-particle $A'$ --then:
    \begin{enumerate}
        \item All variables in $v_{A'}$ except $k_{A'}$, $t_{A'}$, and $t_{A'\,{\rm old}}$, have identical values as those in $v_A$.
            The other variables will be set in the next steps.
        \item The acquired transverse momentum is determined as follows:
            A number $R_{q}\in [0,1]$ is randomly selected from a uniform distribution. 
            Then, the following equation is solved for $q$
            \begin{equation}
                R_{q}=\frac{\int_{q_{\rm min}}^q q'dq'\int_0^{2\pi}\frac{d\varphi_{q'}}{(2\pi)^2} w_A(\mathbf{q}') }{W_A}=\frac{\int_{q_{\rm min}}^q q'dq'\frac{1}{2\pi} w_A(\mathbf{q}') }{\int_{q_{\rm min}}^\infty q'dq'\frac{1}{2\pi} w_A(\mathbf{q}')}\,,
                \label{eq:selq}
            \end{equation}
            where we have implicitly used polar coordinates for the transverse momenta $\mathbf{q}$ ($\mathbf{q}'$), with $q=|\mathbf{q}|$ ($q'=|\mathbf{q}'|$) and $\varphi_q$ ($\varphi_{q'}$) the azimuthal angle in the tranverse plane. 
            In a next step the azimuthal angle $\varphi_{q} \in [0,2\pi]$ is randomly selected from a uniform distribution.
            The transverse momentum transferred to the jet-particle from the medium is obtained as 
            \begin{equation}
            \mathbf{q}=\left(\begin{array}{c}
            q \cos(\varphi_q)  \\
            q \sin(\varphi_q) 
            \end{array}\right)\,.
            \end{equation}
            The new transverse momentum of the particle $A'$ after scattering is obtained as 
            \begin{equation}
            \mathbf{k}_{A'}=\mathbf{k}_A+\mathbf{q}\,.
            \end{equation}
    \item The time of the next scattering or branching $t_{A'}$ is determined by selection of a random variable $R_{t}\in [0,\,1]$ and solving
        \begin{equation}
        R_{t}=\Delta_{typ_A}(x_A,t_{A'}-t_A))\,.
        \label{eq:tsel2}
        \end{equation}
    \item Set $t_{A'\,{\rm old}}=t_A$.
    \item If $t_{A'}\geq t_L$ add $v_{A'}$ to $\mathcal{S}_{\rm fin}$, otherwise add $v_{A'}$ to $\mathcal{S}'$.
    \end{enumerate}
    \item If particle $A$ undergoes a splitting into two particle $B$ and $C$, then:
    \begin{enumerate}
        \item First, the type of splitting is determined: If the particle $A$ is a quark (antiquark) it is presumed that it undergoes a splitting into a quark (antiquark) and a gluon. However, in case particle $A$ is a gluon, the type of reaction is determined as follows: 
        A number $R_{react}\in [0,1]$ is randomly selected from a uniform distribution.
        If 
        \begin{equation}
          R_{react} \leq \frac{\rho_{gg}(x_A)}{\rho_{qg}(x_A)+\rho_{gg}(x_A)} \,,
          \label{eq:selsplittyp}
        \end{equation}
        then the gluon splits into a gluon pair, otherwise into a quark-antiquark pair. Consequently, variables $typ_B$ and $typ_C$ are set accordingly.
        \item The light-cone energy fraction $z$ of particle $B$ with regard to the light cone energy of particle $A$ is selected. 
        To this end, one can define
        \begin{equation}
            \mathcal{K}_{BA}(z):=\frac{1}{2}\sqrt{\frac{f_{BA}(z)}{z(1-z)}}P_{BA}(z)\,,
        \end{equation}
        and note that
        \begin{equation}
            \alpha_s \int_{Q>0} \frac{d^2\mathbf{Q}}{(2\pi)^2}\mathcal{K}_{BA}(\mathbf{Q},z,p_+)=\frac{\alpha_s}{\pi}\sqrt{\frac{\hat q}{N_cp_+}}\mathcal{K}_{BA}(z)\,.
        \end{equation}
        Thus, a number $R_z\in[0,1]$ is selected randomly from a uniform distribution and the equation
        \begin{equation}
            R_z=\frac{\int_\epsilon^z dz'\mathcal{K}_{BA}(z')}{\int_\epsilon^{1-\epsilon} dz'\mathcal{K}_{BA}(z')}\,,
            \label{eq:zsel}
        \end{equation}
        is solved for $z$. The total light cone energy fraction of particle $B$ and $C$ with regard to the initial cascade particle are set as 
        \begin{equation}
            x_B=x_Az\,,
        \end{equation}
        and as
        \begin{equation}
            x_C=x_A(1-z)\,,
        \end{equation}     
        respectively.
        \item Instead of a direct selection from the splitting kernel, the value for $Q^2$ can be more easily obtained, by selecting a value for $u:=\frac{Q^2}{2k_{BR}^2}$, because the kernel $\mathcal{K}_{BA}(\mathbf{Q},z,xp_+)$ given in Eq.~(\ref{eq:Kqz}) factorizes into a distribution of $u$ and $z$. 
        Thus, a number $R_u\in [0,1]$ is randomly selected from a uniform distribution.
        Then, the equation 
        \begin{equation}
            R_u=\frac{\int_0^u du' \sin (u') {\rm e}^{-u'}}{\int_0^\pi du' \sin (u') {\rm e}^{-u'}}\,,
            \label{eq:usel}
        \end{equation}
        is solved.
        Please note that an upper cutoff $u=\pi$ was introduced here, since at higher values of $u$ the splitting kernel would become negative.
        After determination of $u$ the absolute values of the transverse momentum $Q$ is determined as 
        \begin{align}
            Q^2&= 2\sqrt{z(1-z)p_+\frac{\hat{q}}{N_c}f_{BA}(z)}u\,.\\
        \end{align}
        Then, an azimuthal angle $\varphi\in [0,2\pi]$ is randomly selected from a uniform distribution. Then, $\mathbf{Q}$ can be obtained as
        \begin{equation}
            \mathbf{Q}=\left(\begin{array}{c}
                Q\cos (\varphi) \\
                Q\sin (\varphi)   
            \end{array}
            \right)\,.
        \end{equation}
        The transverse momenta of particle $B$ and $C$ are determined as
        \begin{align}
            &\mathbf{k}_B=z\mathbf{k}_A+\mathbf{Q}\,,\\
            &\mathbf{k}_C=(1-z)\mathbf{k}_A-\mathbf{Q}\,.
        \end{align}
        \item For both particles, $B$ and $C$ the respective times of their next scatterings or splittings, $t_B$ and $t_C$ are determined by independently randomly selecting two variables $R_{t_B}\in[0,1]$ and $R_{t_C}\in[0,1]$ from uniform distributions and then solving the equations
        \begin{align}
        R_{t_B}&=\Delta_{typ_B}(x_B,t_{B}-t_A)\,,
        \label{eq:tsel3}
        \\
        R_{t_C}&=\Delta_{typ_C}(x_C,t_{C}-t_A)\,.
        \label{eq:tsel4}
        \end{align}
        \item Set $t_{B\,old}=t_A$ and $t_{C\,old}=t_A$.
        \item With the previous steps the sets of variables $v_B$ and $v_C$ for particles $B$ and $C$, respectively have been determined.
        In a last step it is verified
        \begin{enumerate}
            \item if $t_B\geq t_L$ $v_B$ is added to $\mathcal{S}_{fin}$, otherwise to $\mathcal{S}'$,
            \item if $t_C\geq t_L$ $v_C$ is added to $\mathcal{S}_{fin}$, otherwise to $\mathcal{S}'$.
        \end{enumerate}
    \end{enumerate}
\end{enumerate}

In order to be able to obtain the random selection of variables in the algorithm above, the \tmdice~ program first calculates the following list of probabilities and inverses of cumulative distribution functions:
\begin{itemize}
\item The functions $\rho_{\rm BA}(x)$ and the values $W_A$ ($A=q,g$ and $B=q,g$) allowing for the solutions of Eqs.~(\ref{eq:tsel}), (\ref{eq:tsel2}), (\ref{eq:tsel3}), and (\ref{eq:tsel4}), as well as Eqs.~(\ref{eq:spvssc}) and (\ref{eq:selsplittyp}).
\item The cumulative distribution functions $\frac{\int_{q_{\rm min}}^q q'dq'\int_0^{2\pi}\frac{d\varphi_{q'}}{(2\pi)^2} w_A(\mathbf{q}') }{W_A}$, $\frac{\int_\epsilon^z dz'\mathcal{K}_{BA}(z')}{\int_\epsilon^{1-\epsilon} dz'\mathcal{K}_{BA}(z')}$, and $\frac{\int_0^u du' \sin (u') {\rm e}^{-u'}}{\int_0^\pi du' \sin (u') {\rm e}^{-u'}}$ as functions of $q$, $z$, and $u$, respectively, and the corresponding inverses of these cumulative distribution functions, allowing for the solutions of Eqs.~(\ref{eq:selq}), (\ref{eq:zsel}), and (\ref{eq:usel}).
\end{itemize}

\section{Usage of the program code}
\label{sec3}
\subsection{Installation}
The code can be downloaded from
\begin{verbatim}
https://github.com/Rohrmoser/TMDICE
\end{verbatim}
in the form of a \verb#.zip# file.
In order to be able to execute the code or include it as a library into a C++ program the \verb#.zip# file needs to be expanded (via \verb#unzip#) into a folder, which from now on will be referenced as 
\verb#$TMDICEfolder#.
The main functions of the algorithm can be loaded as a library into C++ code, so in order to be able to run the code within a C++ program the basic requirements are a C++ compiler 
and include the following header files in the program code/compile the program code including the following C++ source files in \verb#$TMDICEfolder#:
\begin{itemize}
    \item "TMDICE.cpp"
    \item "TMDICE.h"
    \item "TMDICE\_lib.cpp"
    \item "TMDICE\_lib.tpp"
    \item "TMDICE\_lib.h"
    \item "deps.h"
\end{itemize}
Thus, in order to include \tmdice\, into a C++ program it is necessary to include the following line at the beginning of the main file of the program,
\begin{verbatim}
#include "TMDICE.h"
\end{verbatim}
and compiling the programs with reference to source files and location of the library files, which in the g++ compiler of the GNU-compiler collection
, will look like this:
\begin{verbatim}
g++ -std=c++11 <main> $TMDICEfolder/TMDICE.cpp \
$TMDICEfolder/TMDICE_lib.cpp -I$TMDICEfolder <flags>
\end{verbatim}
where \verb#<main># are the C++ source files and location of the header files for the main program and \verb#<flags># symbolizes additional compiler flags (unrelated to \tmdice).
The C++ compiler that is used needs to be able to compile code that follows the C++11 standard of the C++ programming language (or also more recent standards). In the example for g++ above this is achieved by the compiler flag  \verb#-std=c++11#. 

Under Linux-systems on which the GNU-compiler collection
is installed it is possible to create a code example for the production of numerous cascades, by typing the following in a terminal window:
\begin{verbatim}
cd $TMDICEfolder
bash make_demo.sh $DEMOfolder
\end{verbatim}
where \verb#$DEMOfolder# is the path to the example.
In order to compile and run the example type
\begin{verbatim}
cd $DEMOfolder
bash makefile.sh
./demo.out <outputfile>
\end{verbatim}
where \verb#<outputfile># is the name of a newly created file into which the output of the Monte-Carlo program is written.
In order to allow to plot data (at least in the form of the corresponding fragmentation functions) from \verb#<outputfile># a Mathematica-notebook "evaluation\_demo.nb" has been added (which can be executed successfully under Mathematica version 12).
\subsection{Initialization of the Monte-Carlo program}
In order to be able to execute the Monte-Carlo generation of jets, it is first of all necessary to set the relevant input-parameters and calculate the necessary probabilities and inverses of the cumulative distribution functions.
The parameters are read into the program by the function \verb#readTMDICEparameters# 
for which there are the following two possibilities:
\begin{enumerate}
    \item 
    \begin{verbatim}
readTMDICEparameters(input_file_name);
    \end{verbatim}
    where \verb#input_file_name# is a string-variable that gives the file name of an input file in which the parameters are listed in the following way: 
    Every line of the file contains only a single parameter. First the name of the parameter is written, then, separated by a space-character the value of the parameter.
    \item \begin{verbatim}
readTMDICEparameters({{name_par_1, value_par_1},{name_par_2, value_par_2},
...,{name_par_n, value_par_n}});
    \end{verbatim}
   where $n$ different parameters are given via their names \verb#name_par_i# and their numerical values \verb#value_par_i# (for $i=1,\,n$).
\end{enumerate}
The order in which the parameters are listed is irrelevant in both of the above possibilities. 
Tab.~\ref{tab:med_params} lists the set of parameters together with the names that allow to address them, the dimensions in which the values need to be given, the possible values and (if available) their default values.
The time-scales for the evolution can be given either directly as $t_0$ and $t_{\rm max}$ in units of [fm/c] or as dimensionless quantities 
\begin{align}
    \tau_0=\frac{t_0}{t^\ast}\,,&&\tau_{\rm max}=\frac{t_{\rm max}}{t^\ast}\,,&&\textrm{with }\frac{1}{t^\ast}=\frac{\alpha_s}{\pi}\sqrt{\frac{\hat{q}}{p_+}}\,.
\end{align}
A certain set of parameters must be specified by the user, in order for the program to run properly.
These parameters are:
\begin{itemize}
    \item the number of colors $N_c$,
    \item the density of scatterers $n_{\rm med}$,
    \item the average squared transverse momentum transfer $\hat{q}$,
    \item the initial light-cone particle energy $p_+$,
    \item either both $t_0$ and $t_{\rm max}$ or both $\tau_0$ and $\tau_{\rm max}$,
    \item either $\alpha_s$ or $\bar{\alpha}_s=\frac{\alpha_s}{\pi}$.
\end{itemize}
If one or more of the necessary parameters in the list above is not specified by the user, the function \texttt{readTMDICEparameters} will give a corresponding error message upon execution of the program and the program will stop. However, it is possible to give both $\alpha_s$ and $\bar{\alpha}_s$ or both sets $t_0$ and $t_{\rm max}$ as well as $\tau_0$ and $\tau_{\rm max}$. In that case \texttt{readTMDICEparameters} will execute without error, but in the next mandatory step for execution of \tmdice~(the function \texttt{setTMDICE}, discussed below) the input for $\bar{\alpha}_s$ or $\tau_0$ and $\tau_{\rm max}$ is ignored and rather values consistent with the respective choices for $\alpha_s$ or $t_0$ and $t_{\rm max}$ are obtained by the program.
If a certain non-mandatory parameter is not given, the program will assume the default value of that parameter. 
There also have been implemented the flags \verb#scat# and \verb#ktsplit#, which specify the type of scattering kernel and, respectively, whether the branchings are considered collinear (i.e.: in every branching for the produced particles only the momentum fractions are determined, while there is no transfer of transverse momentum from the medium during branching) or non-collinear.
\begin{table}[h!]
    \centering
    \begin{tabular}{p{7.5cm}|c|c|p{3.8cm}|c}
         variable description& variable name&Dimension& Possible  &default \\
         & in program&&values&value\\\hline
         initial particle momentum fraction $x_1$&\verb#x1#&[1]&value in range $[0,1]$&$1$\\
         initial particle transverse momentum$||\mathbf{k}_1||$&\verb#kt1#&[GeV]&positive real value&$0$\\
         initial particle type $typ_1$&\verb#typ1#&[1]&$1$ for quarks,$-1$ for antiquarks, $2$ for gluons&$2$\\
         initial particle light-cone energy $p_+$ & \verb#emax#&[GeV]&positive real value&--\\ 
         number of colors $N_c$&\verb#nc#&[1]&positive integer value&$3$\\
         density of scatteres $n_{\rm med}$& \verb#ndens#& [GeV$^3$]&positive real value& --\\
         average squared transverse momentum transfer $\hat{q}$& \verb#qhat#&[GeV$^2$/fm]&positive real value&--\\
         time of start of jet-evolution $t_0$&\verb#tmin#&[fm/c]&positive real value&--\\
         time of end of jet-evolution $t_{\rm max}$&\verb#tmax#&[fm/c]&positive real value&--\\
         $\tau_0=\frac{t_0}{t^\ast}$&\verb#taumin#&[1]&positive real value&--\\
         $\tau_{\rm max}=\frac{t_{\rm max}}{t^\ast}$&\verb#taumax#&[1]&positive real value&--\\
         $\alpha_s$&\verb#alphas#&[1]&positive real value&--\\
         $\bar{\alpha}_s=\frac{\alpha_s}{\pi}$&\verb#alphabar#&[1]&positive real value&--\\
         $\epsilon$&\verb#xeps#&[1]&value in range $[0,1]$&$10^{-4}$\\
         $x_{\rm min}$&\verb#xmin#&[1]&value in range $[0,1]$&$10^{-4}$\\
         $q_{\rm min}$&\verb#qmin#&[GeV]&positive real value&$0.1$\\
         medium temperature $T$&\verb#T#&[GeV]&positive real value&--\\
         Debye mass $m_D$& \verb#md#&[GeV]&positive real value&--\\
         type of scattering kernel &\verb#scat#&[1]&$0\dots$no scattering, i.e.: $w_g(\mathbf{q})=0$&$0$\\&&& $1\dots w_g$ given in Eq.~(\ref{eq:wq1})
         \\&&& $2\dots w_g$ given in Eq.~(\ref{eq:wq2})
         \\&&& $3\dots w_g$ given in Eq.~(\ref{eq:wq3})&
         \\
         collinear or non-collinear splitting&\verb#ktsplit#&$[1]$&$0\dots$collinear splitting,&$1$\\
         &&&$1\dots$non-collinear splitting\\
     \end{tabular}
    \caption{List of parameters for the program.}
    \label{tab:med_params}
\end{table}

In a next step, the parameters of the program are fixed to the previously given values (or the default values) and the probabilities and inverses of the cumulative distribution functions for the in-medium jet evolution are calculated in the function 
\begin{verbatim}
setTMDICE();
\end{verbatim}
This step is mandatory in order to be able to obtain results for in-medium jet-evolution later on.
Please note that the calculation of all partition functions may take a lot of time, however, in the present version of the program, \verb#setTMDICE()# needs to be executed only once\footnote{For the purpose of changing $E_{\rm max}$, $t_{\rm max}$, $t_0$, $x_1$, $k_1$, $typ_1$ later on, the corresponding setter functions \verb|setEmax(ee);|, \verb|settmax(tup);|, \verb|settmin(tmin);|, \verb|setx1(xx);|, \verb|setkt1(kk);|, and \verb|settyp1(ttyp);| were created (where \verb|ee|, \verb|tup|, \verb|tmin|, \verb|xx|, \verb|kk|, and \verb|ttyp| are the numerical values of $E_{\rm max}$, $t_{\rm max}$, $t_0$, $x_1$, $||\mathbf{k}_1||$, and $typ_1$ respectively).}
\subsection{Execution of the Monte-Carlo algorithm}
In \tmdice\, jets are defined as objects of a class \verb#TMDICEevent#, that contains the variables of the jet-particles together with functions that allow to generate a jet via the Monte-Carlo algorithm. 
Thus, in order to execute the Monte-Carlo generation of a single jet it is necessary to first define an instance of the \verb#TMDICEevent# class, which will be labeled here as \verb#jet#, in the following way:
\begin{verbatim}
TMDICEevent jet;
\end{verbatim}
To execute the Monte-Carlo algorithm, the class-function \verb#make_event# needs to be executed (here for the example of a \verb#TMDICEevent# named "jet") as
\begin{verbatim}
jet.make_event();
\end{verbatim}

\subsection{Accessing the produced events}
The produced parton cascades are stored inside the class \verb#TMDICEevent# within the class elements \verb#casc# and \verb#genfin#. The difference is that while \verb#casc# contains all particles of a cascade that evolved between $t_0$ and $t_{\rm max}$, \verb#genfin# only contains those particles that exist at time $t_{max}$ (i.e. those that are emitted at times smaller than $t_{\rm max}$ and are annihilated at times larger than $t_{max}$).
Both, \verb#casc# and \verb#genfin# are defined as vectors of 
sets of particle variables. 
For each particle the following variables are defined

\begin{tabular}{p{5cm}|c|c|p{5cm}}
     variable description& variable name & variable type  &possible\\
              & in program    &            &values    \\\hline
     momentum fraction x& \verb#x#& double& fraction in range $[0,1]$\\
     time of emission & \verb#t_old#&double& value in [fm/c]\\
     time of annihilation & \verb#t#&double& value in [fm/c]\\
     transverse momentum & \verb#kt#&double& value in [GeV]\\
     azimuthal angle of transverse momentum & \verb#phik# &double& value in range $[0,2\pi]$\\
     type of particle & \verb#typ#& double& $1$ for quarks, $-1$ for antiquarks, $2$ for gluons\\
     is $x\leq x_{min}$& \verb#dump#& bool& \verb#true# for $x\leq x_{\rm min}$, \verb#false# otherwise\\
\end{tabular}

The variable \verb#typ# gives $2$ for gluons, $1$ for quarks, $-1$ for antiquarks, and is undefined otherwise.
The variable \verb#dump# gives \verb#false# for particles where $x>x_{min}$ and \verb#true# for particles below the threshold, i.e. $x\leq x_{\rm min}$. Note that if $x\leq x_{\rm min}$ no further parton splitting or scattering is considered by the program, but this particle is nevertheless stored.

To make the notation evident, consider the example that a \verb#TMDICEevent# \verb#jet# has been created, that contains a total number of $N$ particles that exist at time $t_{\rm max}$\footnote{$N$ can be obtained as \verb+N=jet.genfin.size();+}. Then for a particle $i$ ($0\leq i<N$) the momentum fraction $x$ can be obtained as
\begin{verbatim}
jet.genfin.at(i).x;
\end{verbatim}
Analogously,  for one of the $M$ ($M\geq N$) particles that are annihilated at time $t\leq t_{\rm max}$ -- e.g. for particle $j$ ($0\leq j<M$) -- the time of annihilation $t$ can be obtained as 
\begin{verbatim}
jet.casc.at(j).t;
\end{verbatim}

\subsection{Program example}
This is an example of a C++ program code that generates a total number of $10^4$ parton cascades named "jet"  via \tmdice\, and writes the output into a file, whose name is given as a parameter to the program upon its execution
\begin{verbatim}
#include<fstream>
#include "TMDICE.h"

int main(int argc, char **argv)
{
	readTMDICEparameters({ {"nc",3},{"ndens",0.3},{"qhat",1.},{"emax",100.},{"tmin",0.},
	{"tmax",1.},{"alphas",0.3},{"scat",1}});
	
	setTMDICE();
	
	ofstream o;
	o.open(argv[1]);
	for(int i=0;i<pow(10,4);i++)
	{
		TMDICEevent jet;
		jet.make_event();
		
		for(int j=0;j<jet.genfin.size();j++)
		{
			if(jet.genfin.at(j).dump==false){o<<i<<" "<<jet.genfin.at(j).x<<" "
			<<jet.genfin.at(j).kt<<" "<<jet.genfin.at(j).phik<<" "<<jet.genfin.at(j).typ<<endl;}
		}
	}
	o.close();
	
	cout<<"Output written to file: "<<argv[1]<<endl;
}
\end{verbatim}

%\section{Illustrative examples}
\section{Accuracy and example results} 
\label{sec_ex}
In order to show what kind of results can be produced with the program, this section gives some examples.
Furthermore, this section will 
discuss the accuracy of the \tmdice~algorithm.
To this end, one can distinguish between model-dependent, physical influences on accuracy and technical influences on accuracy.
The main goal of the \tmdice~program is to describe the contributions to jet-fragmentation from coherent medium-induced radiations and scatterings -- as given by Eqs.~(\ref{eq:Kqz}) and (\ref{eq:scatkerneldef}), respectively -- in a way that yields parton fragmentation functions, which follow Eqs.~(\ref{eq:BDIM_g}) and (\ref{eq:BDIM_q}).
This approach is a priori limited to some approximations and simplifications:
\begin{itemize}
\item The assumption of a time-independent medium. Therefore, $\hat{q}$, $m_D$, $n_{\rm med}$, and $T$ are considered as constants. The splitting kernel in Eq.~(\ref{eq:Kqz}) has been obtained in~\cite{Blaizot:2012fh,Blaizot:2013vha,Blanco:2021usa} via the assumption of a time-independent, infinite medium. Adopting time-dependent splitting kernels and medium parameters into the \tmdice~program will be considered for future versions of the program.
\item The assumption of a constant coupling constant $\alpha_S$.
\item The possible double counting of emissions. While these processes are suppressed exponentially, due to the behavior of the Sudakov-factor in Eq.~(\ref{eq:sud}) evolution equations as the ones Eqs.~(\ref{eq:BDIM_g}) and (\ref{eq:BDIM_q}) -- and, thus, the \tmdice~algorithm that reproduces such a behavior -- allow for the possibility that two successive emissions happen, where the second emission occurs within the emission time of the first.
\item The neglect of possible additional processes and effects. In the present form, the \tmdice~algorithm describes jet-fragmentation by multiple processes of scatterings and coherent medium induced emissions of single partons. Higher order processes, such as, e.g., the coherent emissions of multiple partons, where the interference effects between the emissions have been included, have not been included. Also effects of color-coherence between successive emissions have been neglected so far.
\end{itemize}

\begin{figure}
 %    \centering
      \includegraphics[scale=0.4]{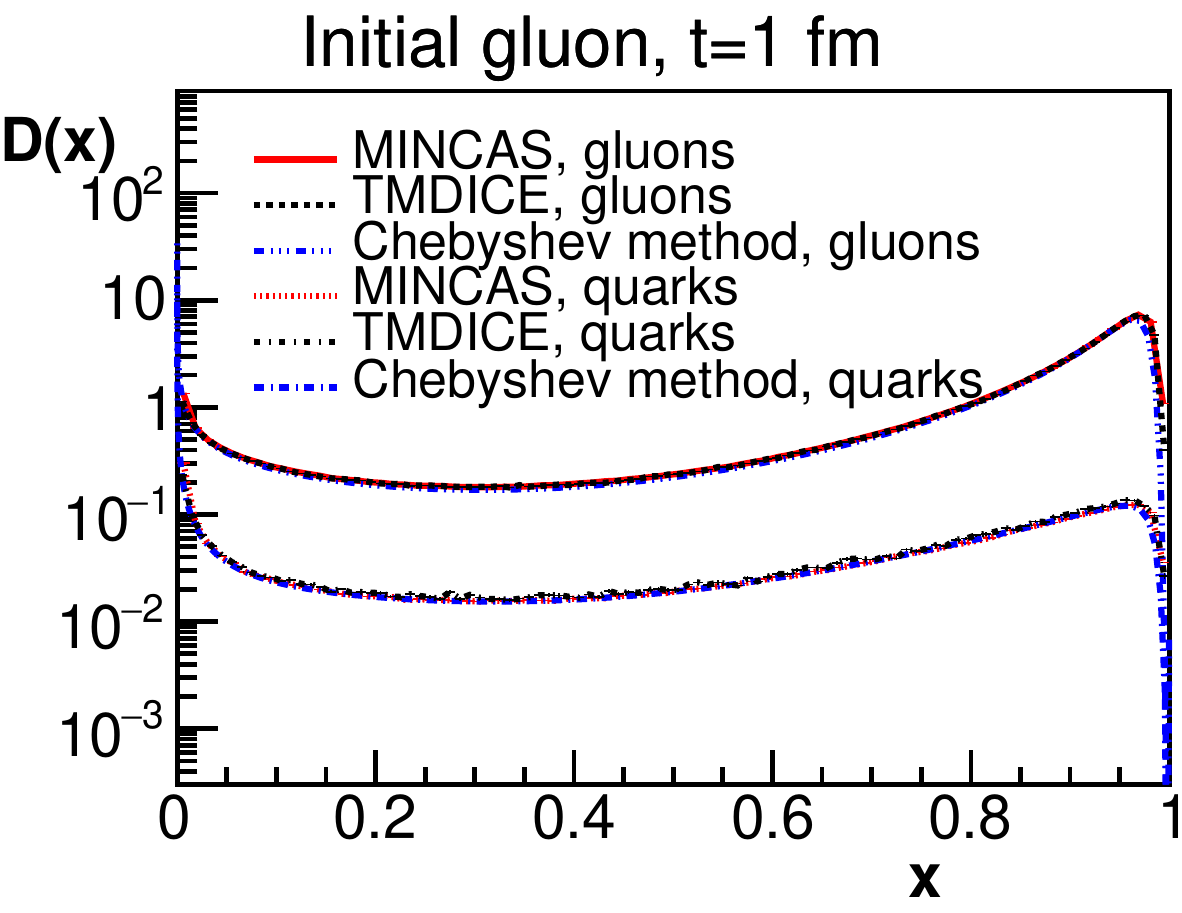}
     \includegraphics[scale=0.4]{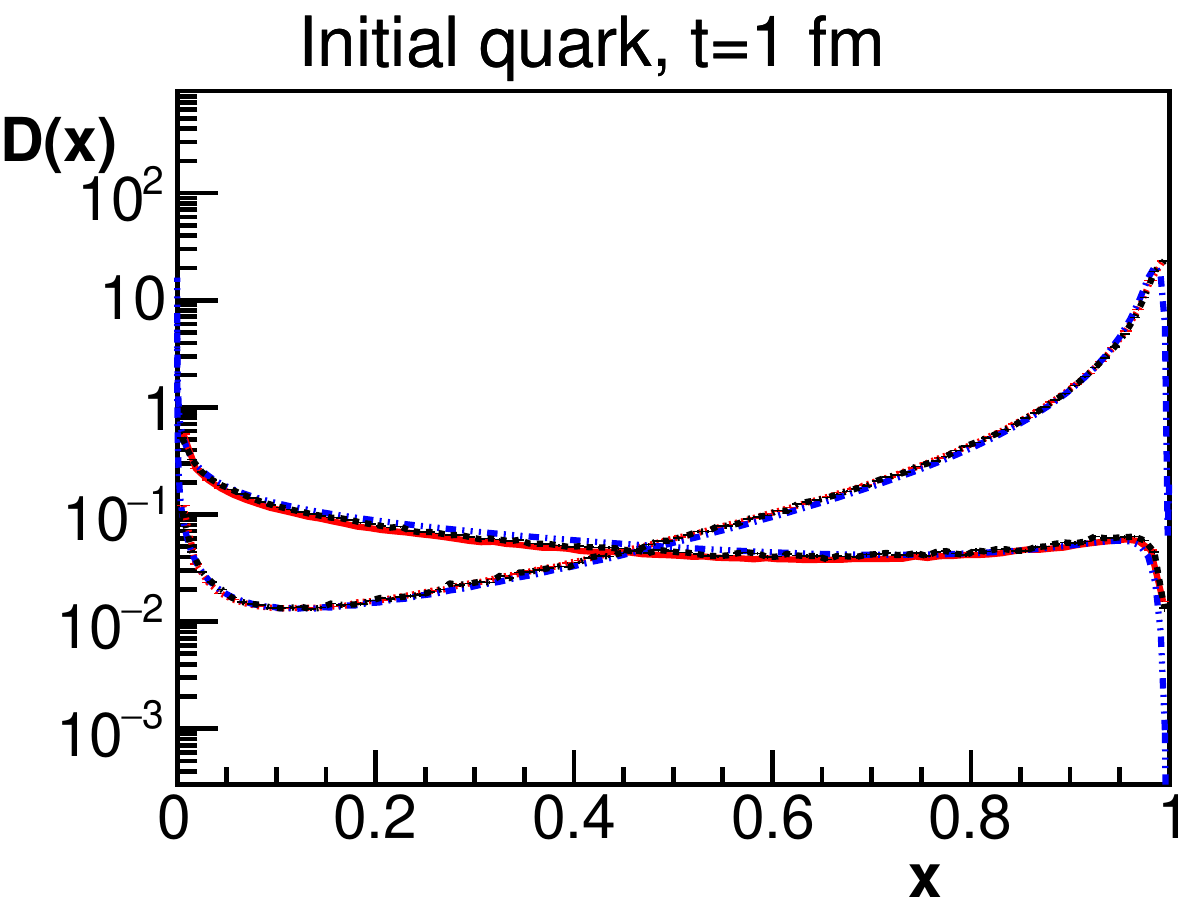}\\
     \includegraphics[scale=0.4]{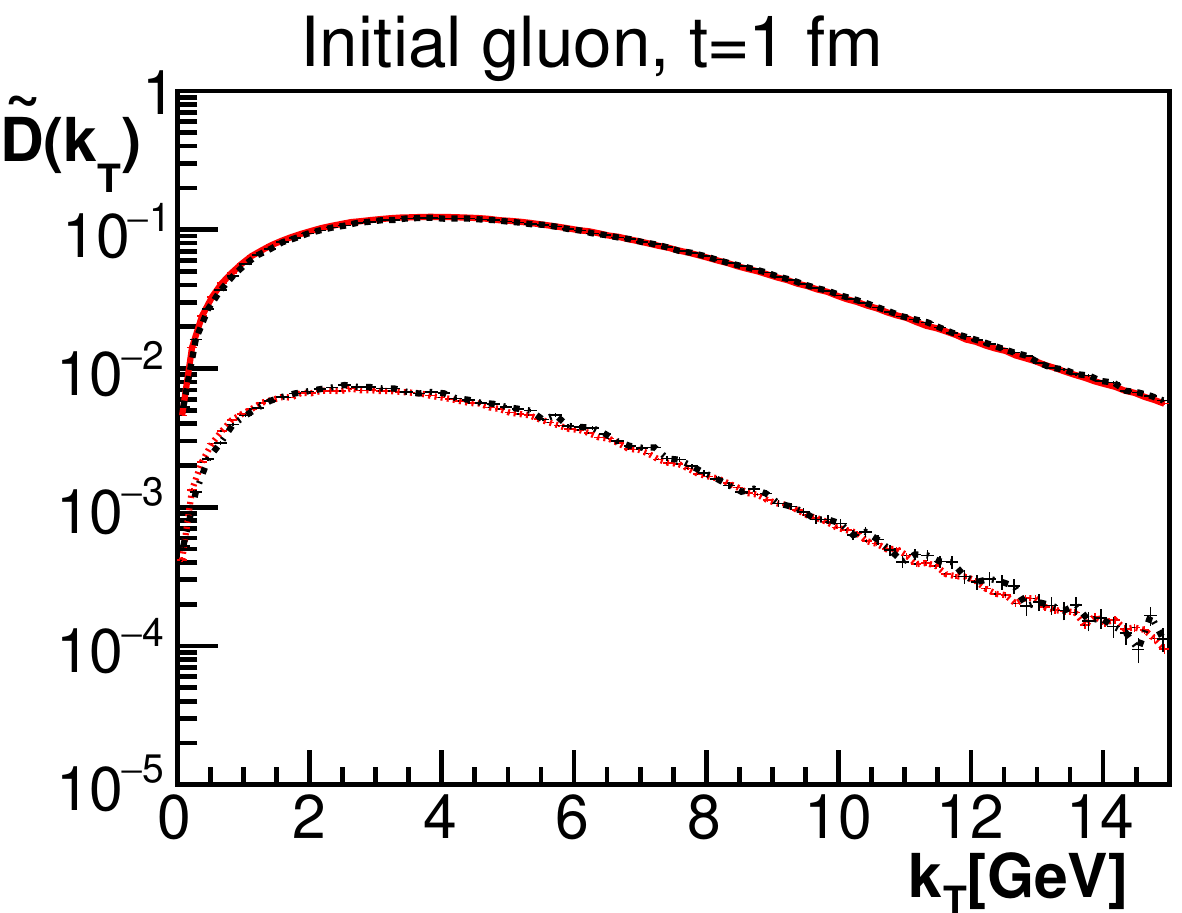}
     \includegraphics[scale=0.4]{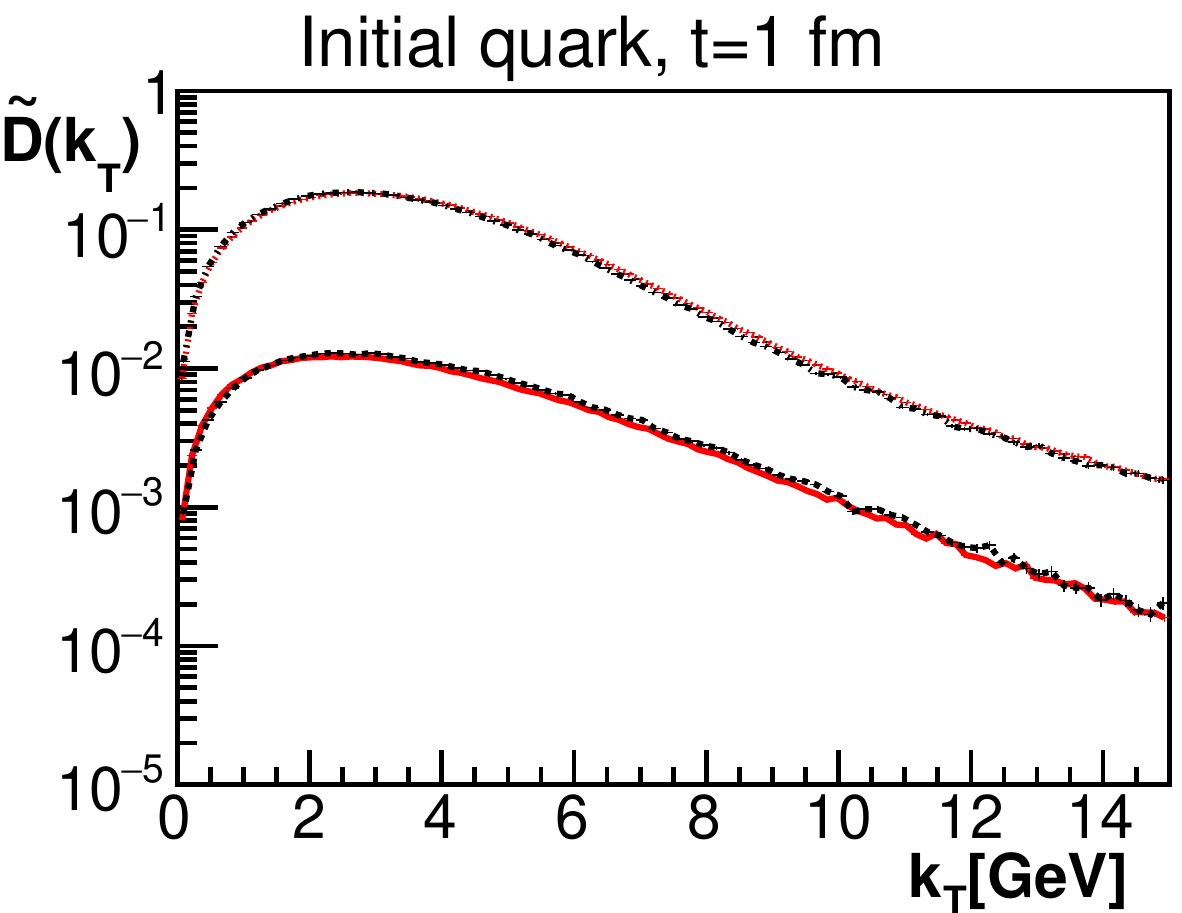}
    \caption{Fragmentation functions $D(x)$ (top) and $\tilde{D}(k_T)$ (bottom) for jet partons initiated by a gluon (left) or a quark (right).}
     \label{fig1}
 \end{figure}

\begin{figure}
\includegraphics[scale=0.4]{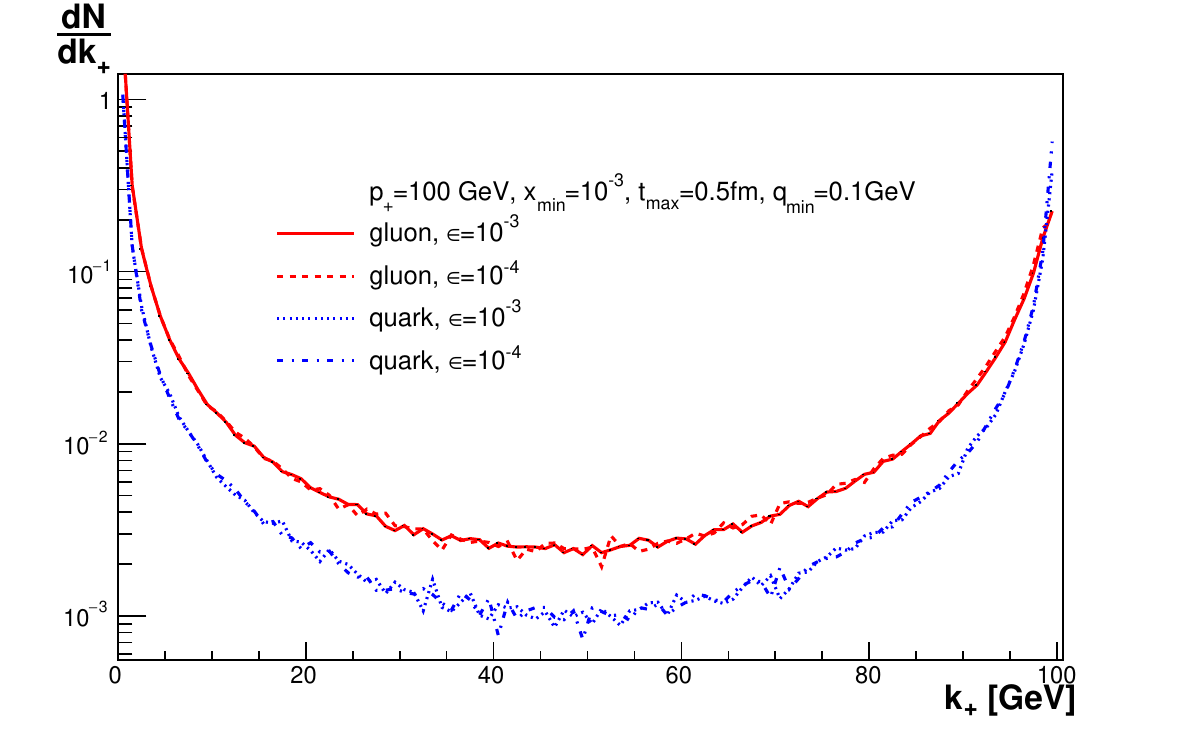}
\includegraphics[scale=0.4]{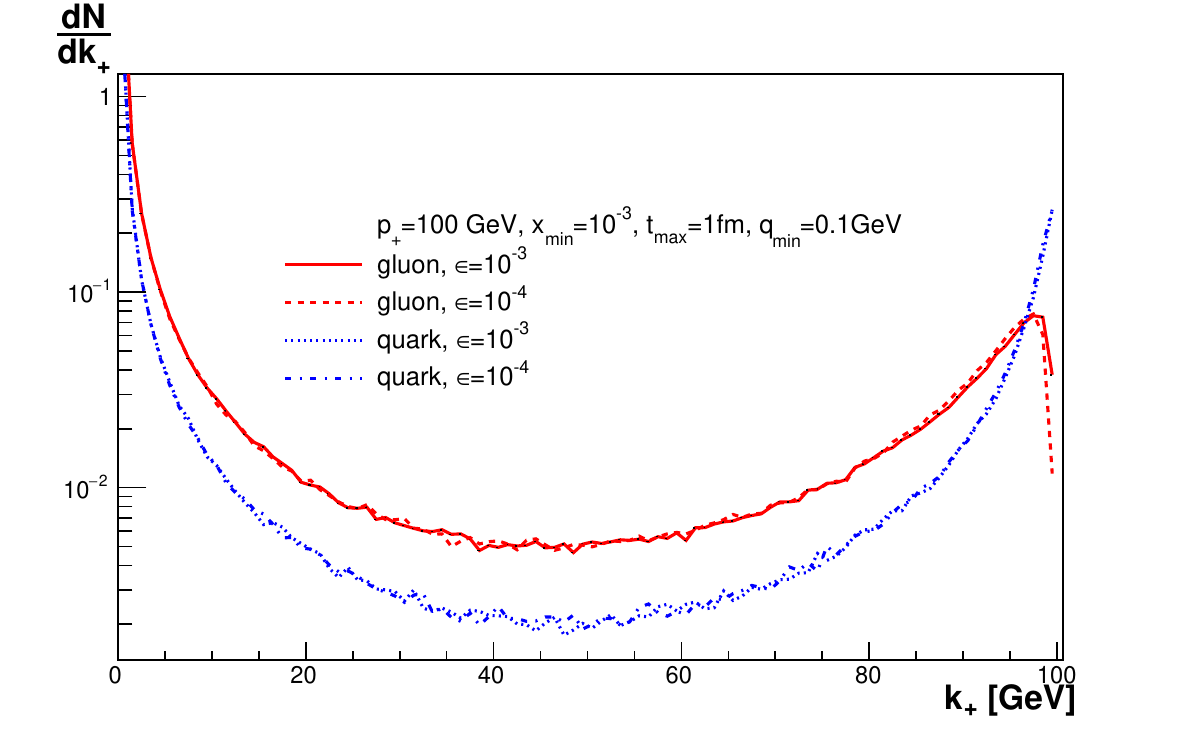}\\
\includegraphics[scale=0.4]{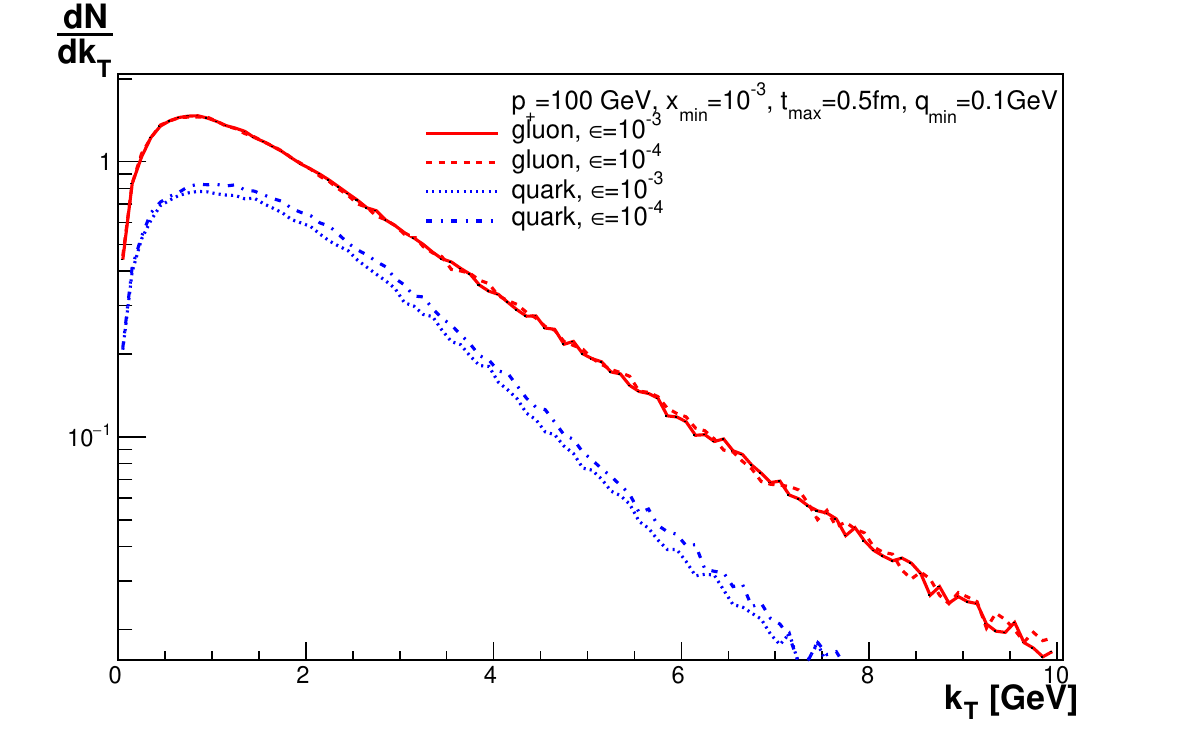}
\includegraphics[scale=0.4]{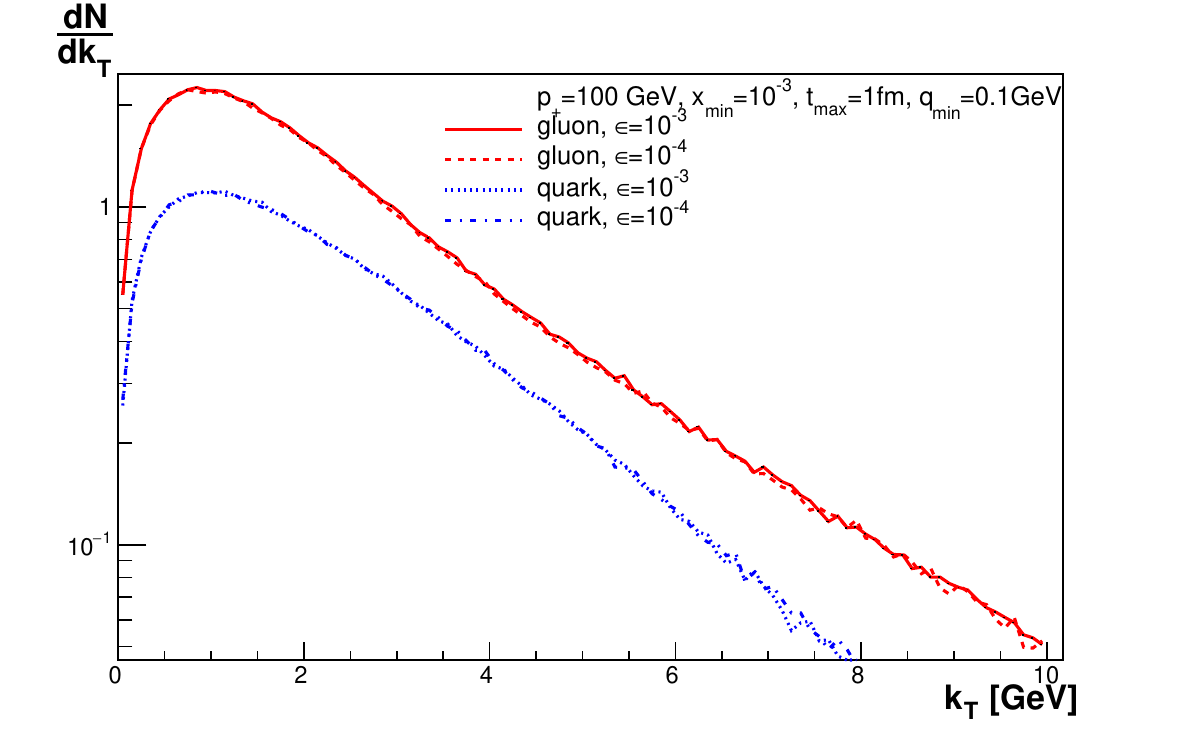}
\caption{Multiplicity distributions in $k_+$ (top) and $k_T$ (bottom) for values for \texttt{tmax} ($t_{\rm max}$) of $0.5$ (left) and $1$ (right) for cascades initiated by quarks and gluons for different values for \texttt{xeps} ($\epsilon$) as indicated. Values for \texttt{xmin} ($x_{\rm min}$) and \texttt{qmin} ($q_{\rm min}$) are $10^{-3}$ and $0.1$, respectively.}
\label{fig2}
\end{figure}

\begin{figure}
\includegraphics[scale=0.4]{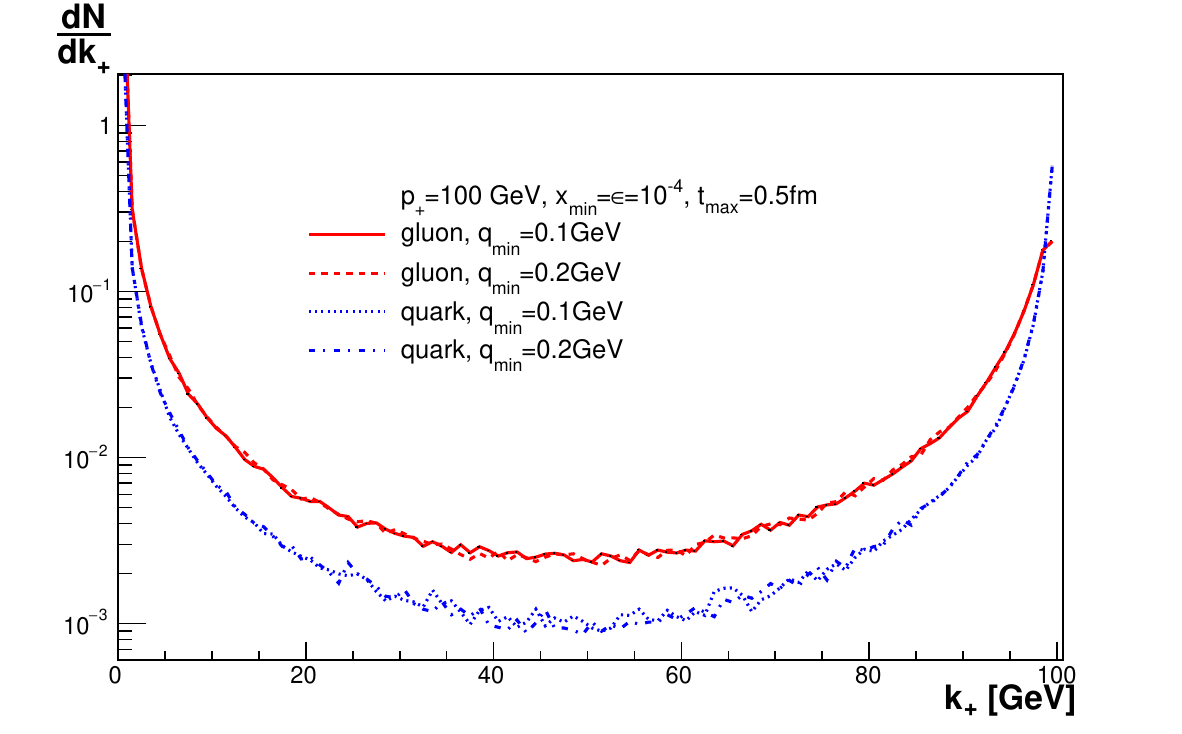}
\includegraphics[scale=0.4]{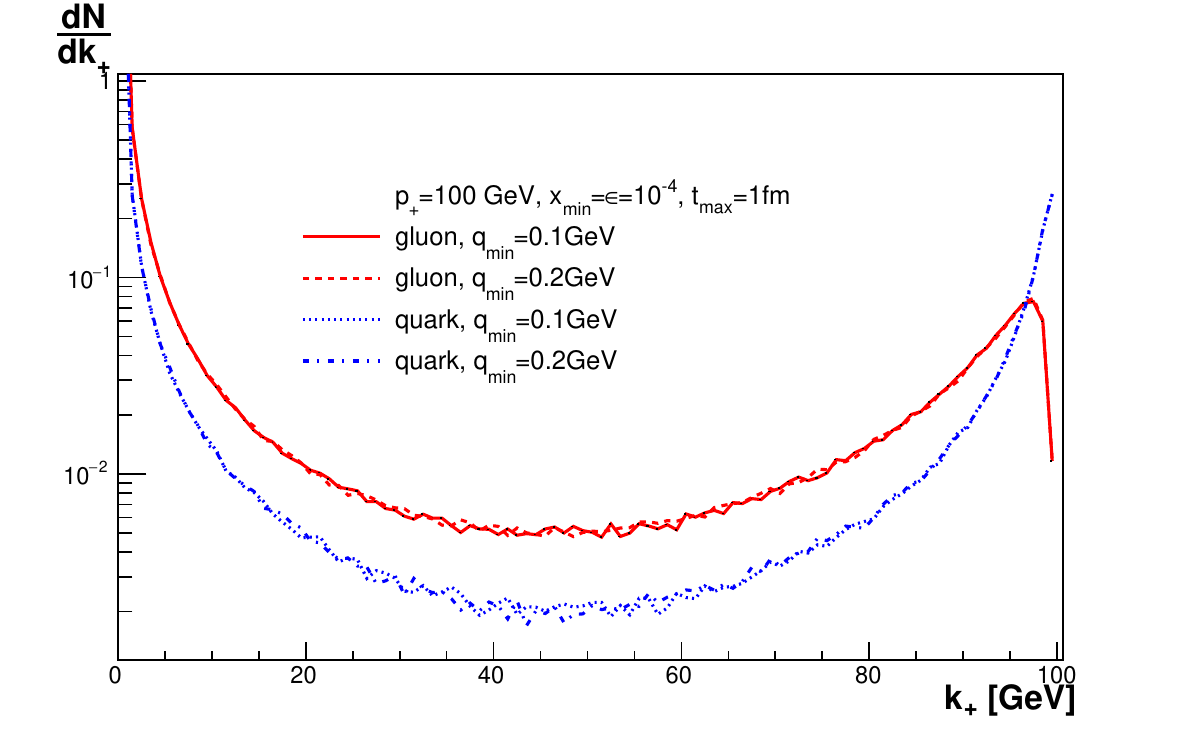}\\
\includegraphics[scale=0.4]{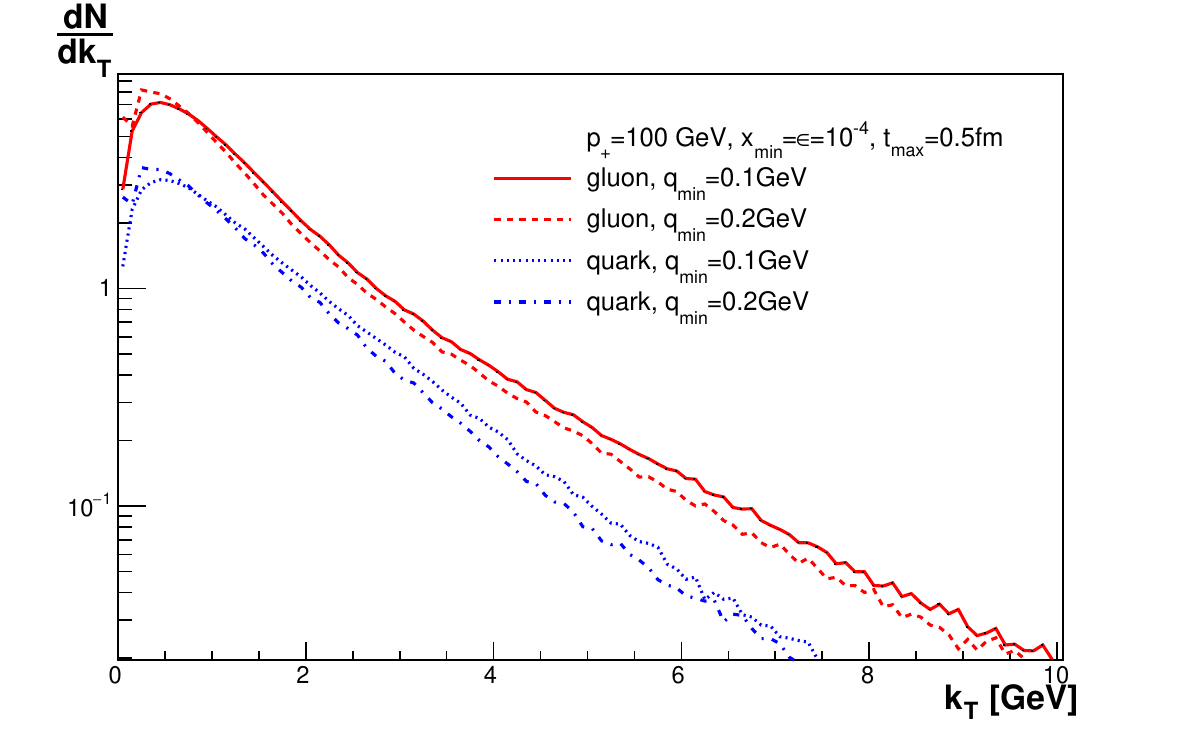}
\includegraphics[scale=0.4]{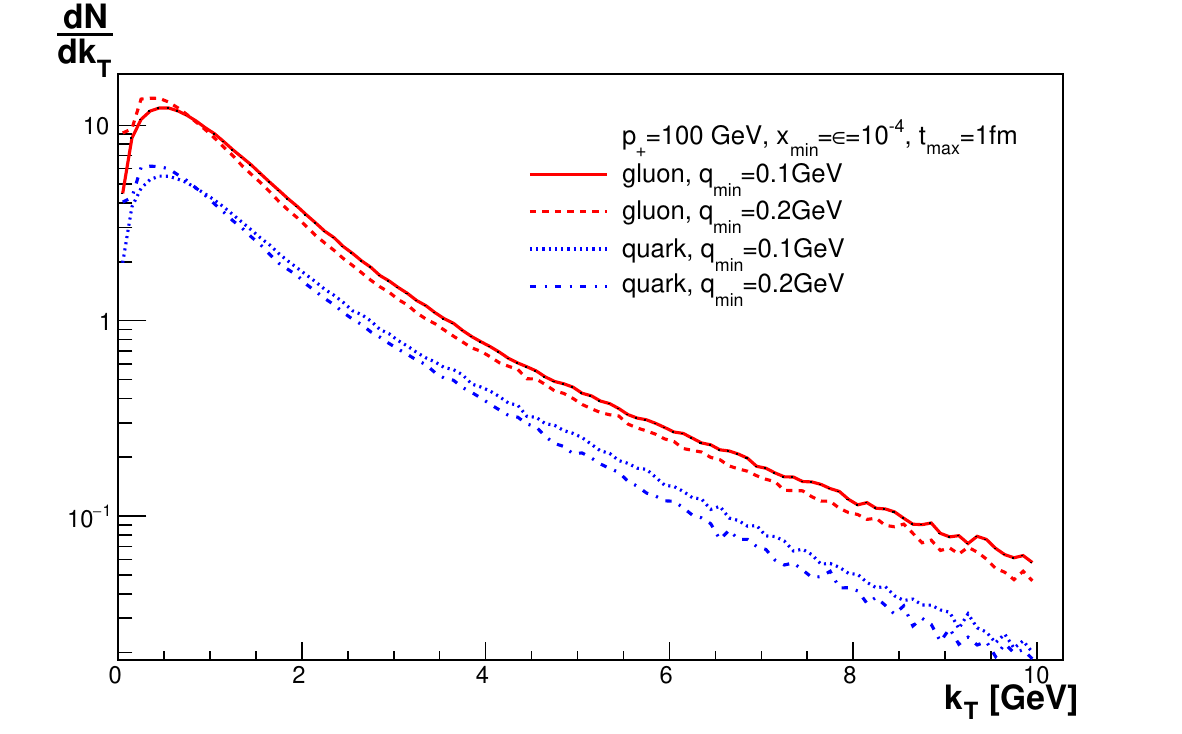}
\caption{Multiplicity distributions in $k_+$ (top) and $k_T$ (bottom) for values for \texttt{tmax} ($t_{\rm max}$) of $0.5$ (left) and $1$ (right) for cascades initiated by quarks and gluons for different values for \texttt{qmin} ($q_{\rm min}$) as indicated. Values for \texttt{xmin} ($x_{\rm min}$) and \texttt{xeps} ($\epsilon$) are $10^{-4}$.}
\label{fig3}
\end{figure}

\begin{figure}
\includegraphics[scale=0.4]{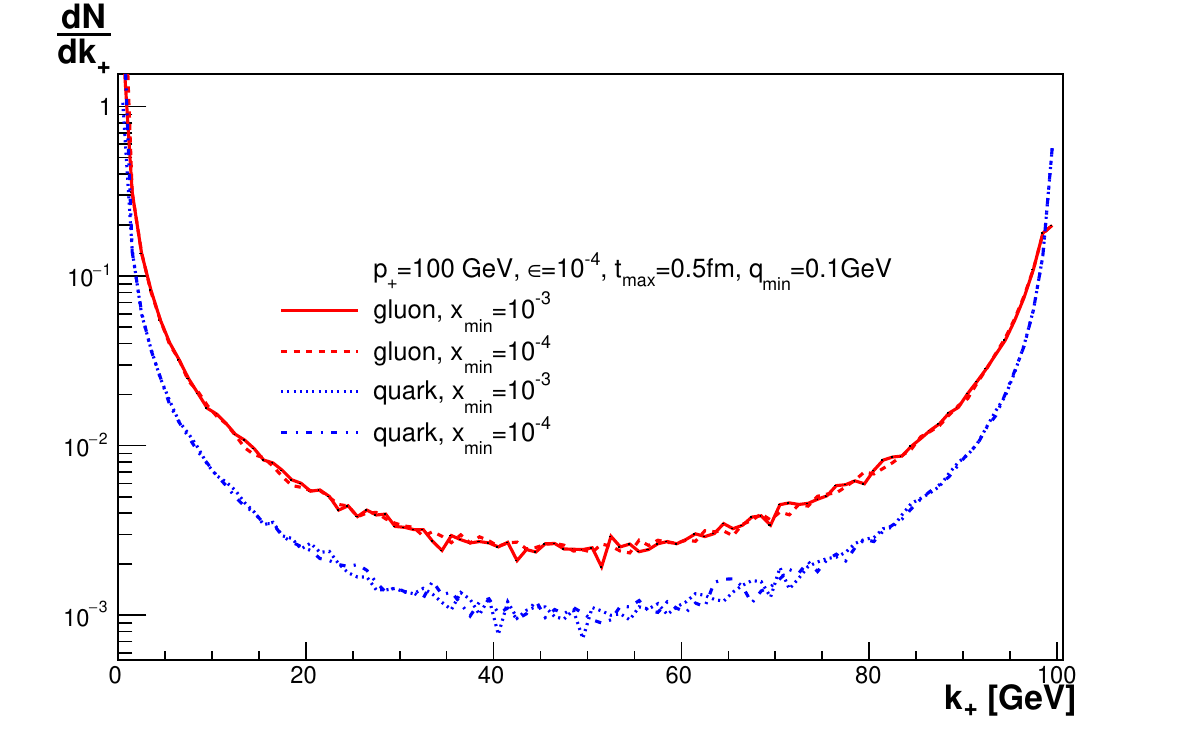}
\includegraphics[scale=0.4]{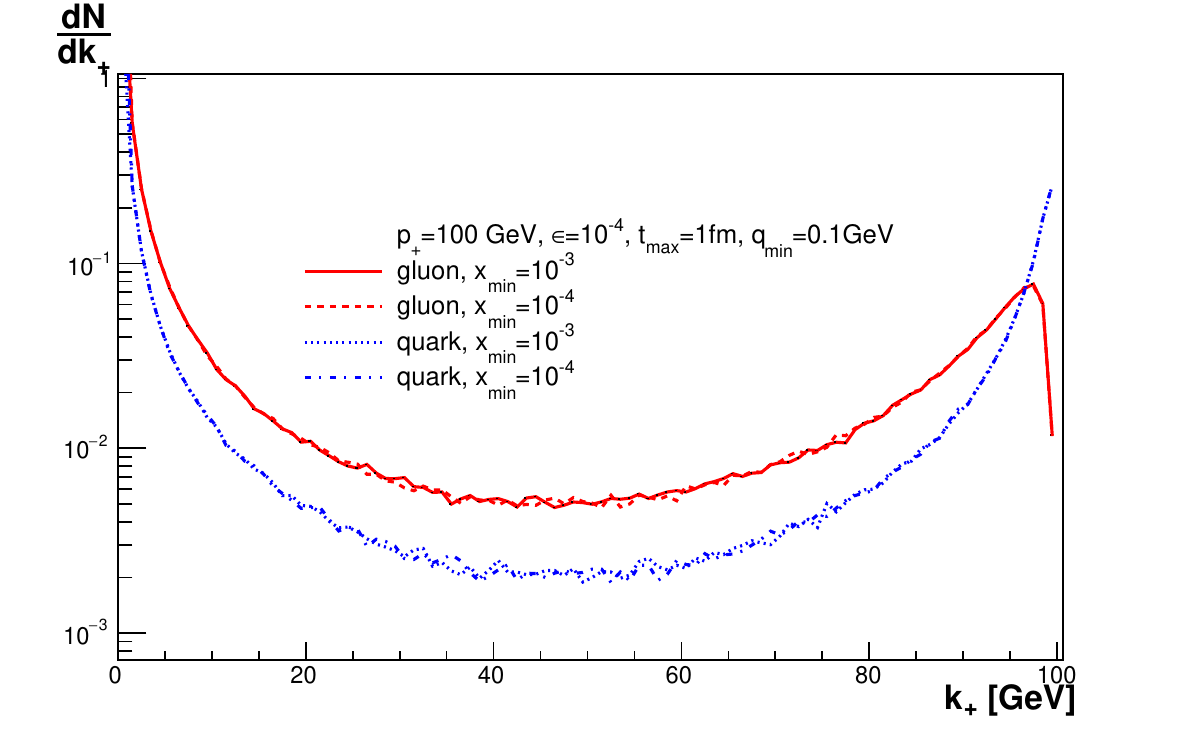}\\
\includegraphics[scale=0.4]{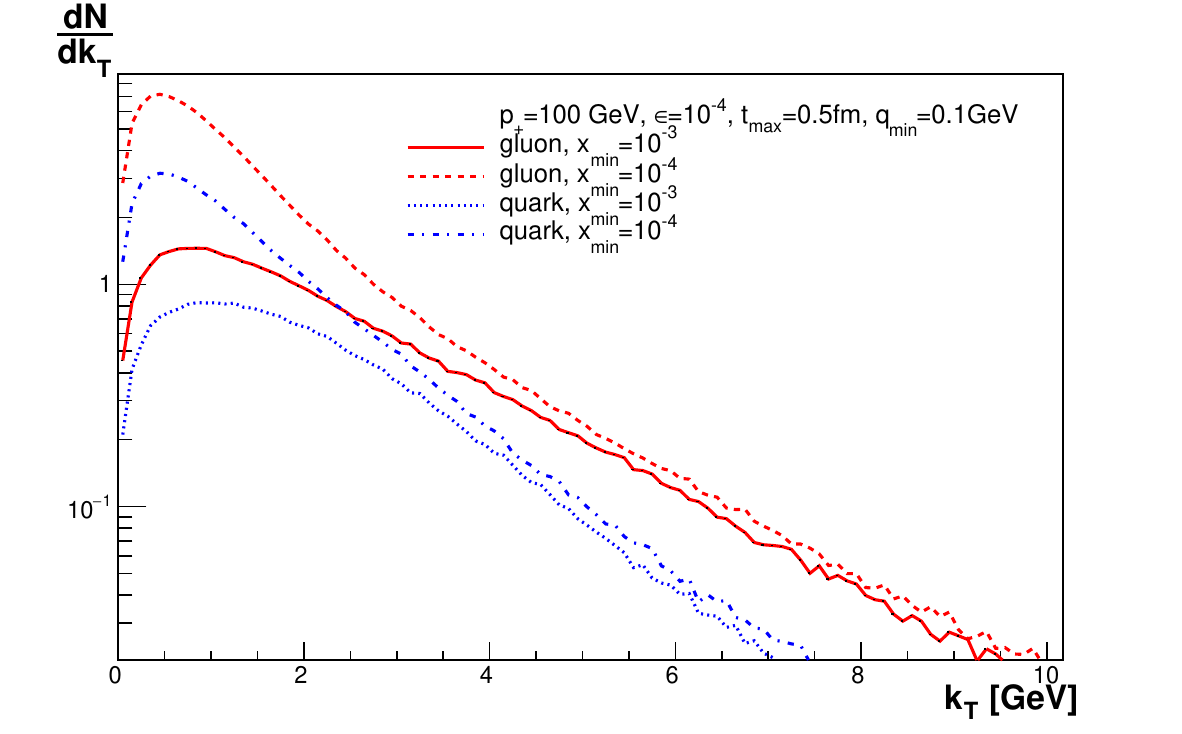}
\includegraphics[scale=0.4]{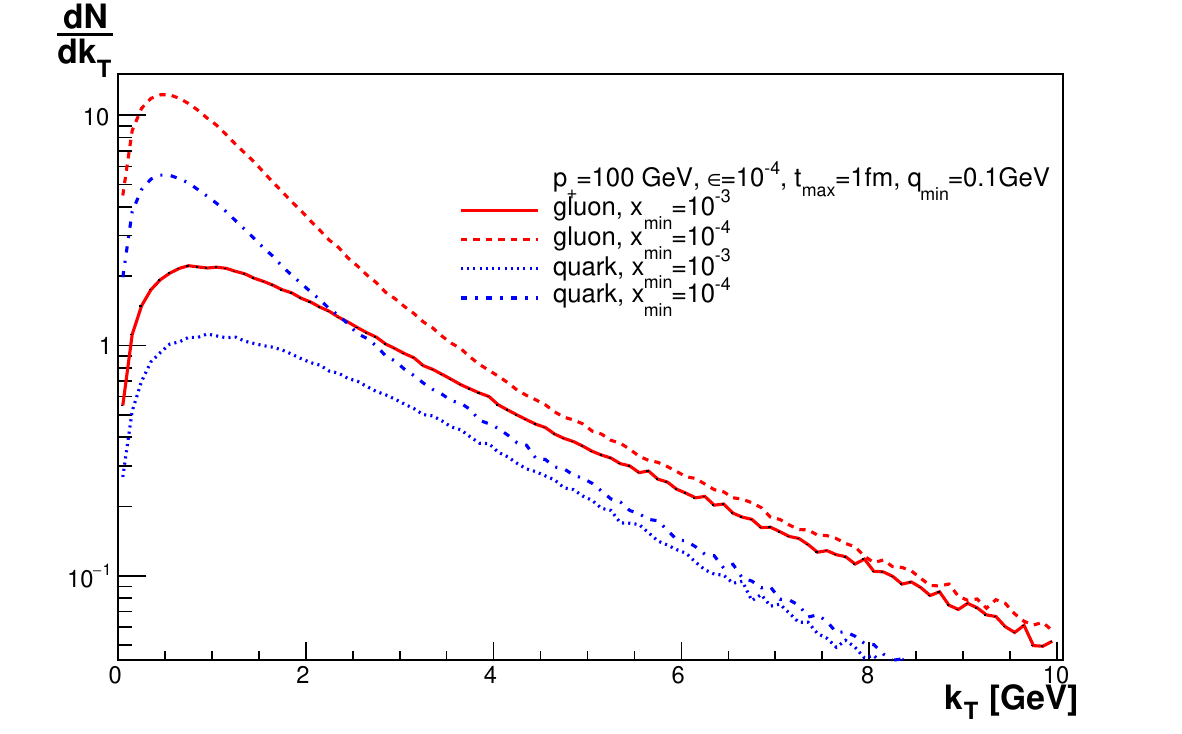}
\caption{Multiplicity distributions in $k_+$ (top) and $k_T$ (bottom) for values for \texttt{tmax} ($t_{\rm max}$) of $0.5$ (left) and $1$ (right) for cascades initiated by quarks and gluons for different values for \texttt{xmin} ($x_{\rm min}$) as indicated. Values for \texttt{xeps} ($\epsilon$) and \texttt{qmin} ($q_{\rm min}$) are $10^{-3}$ and $0.1$, respectively.}
\label{fig4}
\end{figure}

In order to verify that the \tmdice~program describes a jet-evolution, where the parton fragmentation functions follow Eqs.~(\ref{eq:BDIM_g}) and (\ref{eq:BDIM_q}), numerically accurately these fragmentation functions where obtained from \tmdice~via use of Eq.~(\ref{eq:frag_def}) and compared to independent solutions for Eqs.~(\ref{eq:BDIM_g}) and (\ref{eq:BDIM_q}).
To this end first cascades initiated either by a gluon or a quark were simulated for the following set of parameters:

\begin{tabular}{c|c|c|c|c|c|c|c|c|c|c|c|c}
   parameter  &\verb#nc#&\verb#ndens#&\verb#emax#&\verb#qhat#&\verb#tmin#&\verb#tmax#&\verb#alphas#&\verb#scat#&\verb#ktsplit#&\verb#md#\\
    value & $3$&$0.243$&$100$&$1$&$0$&$1$&$0.31415926$&$2$&$1$&$0.993$
\end{tabular}
\\
and the default values for \verb#xmin#, \verb#qmin#, \verb#xeps#, \verb#x1#, and \verb#kt1# from Tab.~\ref{tab:med_params}. The values of $type1$ are set to $1$ and $2$ for the quark and gluon initiated jets respectively. 
Fig.~\ref{fig1} shows results for the fragmentation functions of quarks and gluons integrated over transverse momentum $\mathbf{k}$, 
\begin{align}
   D(x,t_{\rm max})&=\int_{k\geq 0} d\mathbf{k}D(x,\mathbf{k},t_{\rm max})\,, 
\end{align}
and momentum fraction $x$, 
\begin{align}
   \tilde{D}(k_T,t_{\rm max})&:=\int_{x_{\rm min}}^1 dx  \int_0^{2\pi} d\phi_k k_T D(x,\mathbf{k},t_{\rm max})\,, 
\end{align}
where $k_T=||\mathbf{k}||$ and $\phi_k$ is the azimuthal angle of $\mathbf{k}$.
The results are compared to results obtained from the \mincas~Monte-Carlo algorithm~\cite{Blanco:2021usa,Kutak:2018dim} and the Chebyshev method described in~\cite{Blanco:2021usa}. 
\mincas~is a Monte-Carlo algorithm that directly obtains  samples for the fragmentation functions $D(x,\mathbf{k},t_{\rm max})$ for different time-scales $t_{\rm max}$ that follow the evolution equations~(\ref{eq:BDIM_g}) and~(\ref{eq:BDIM_q}). The Chebyshev method expands the fragmentation functions at any given time $t$ on a basis of Chebyshev polynomials. The evolution of the fragmentation functions is obtained by direct solution of the integro-differential evolution equations~(\ref{eq:BDIM_g}) and~(\ref{eq:BDIM_q}): For fragmentation functions given at time $t$ the integrals on the right hand side of Eqs.~(\ref{eq:BDIM_g}) and~(\ref{eq:BDIM_q}) are obtained via numerical integration. Fragmentation functions at later times are obtained via the Euler-method for the solution of differential equations. As an initial condition at time $t_0=0$ a narrow Gaussian distribution with mean value at $x=1$ and standard deviation of $\sigma=10^{-2}$ is assumed instead of a Dirac-distribution, since the latter cannot be suitably well expanded on the basis of Chebyshev polynomials. However, it was so far only possible to obtain results for the evolution of $D(x,t_{\rm max})$. As can be seen in Fig.~\ref{fig1} all three methods agree reasonably well with each other for the evolution of $D(x,t_{\rm max})$ and for $\tilde{D}(k_T,t_{\rm max})$ the results of \tmdice~and \mincas~agree as well.
% %
The main technical influence on the accuracy of results obtained with the \tmdice~algorithm is the number of simulated parton cascades. 
In general, the related statistical error decreases with the number of simulated parton cascades, which needs to be set by the program user.
In order to further test the numerical stability of \tmdice, some of its results are compared for different values of the parameters \verb#qmin#, \verb#xeps#, and \verb#xmin#. 
First, cascades initiated by a gluon (with \verb#type1# set to $2$) and a quark (with \verb#type1# set to $1$) were obtained for the following set of parameters:

\begin{tabular}{c|c|c|c|c|c|c|c|c|c|c|}
   parameter  &\verb#nc#&\verb#ndens#&\verb#emax#&\verb#qhat#&\verb#tmin#&\verb#alphas#&\verb#scat#&\verb#ktsplit# \\
    value & $3$&$0.243$&$100$&$1$&$0$&$0.31415926$&$1$&$0$
\end{tabular}

\noindent
together with the default values for \verb#x1#, and \verb#kt1# from Tab.~\ref{tab:med_params}. The results were obtained for the values of $0.5$ as well as $1$ for \verb#tmax#. 
The scattering kernel in the form of Eq.~(\ref{eq:wq1}) is chosen, because it yields larger transverse momentum broadening effects than the ones of Eqs.~(\ref{eq:wq3}) and~(\ref{eq:wq2}) (as has been demonstrated in~\cite{Blanco:2020uzy} for the case of gluons).
The case of a collinear splitting has been chosen, because the momentum cutoff \verb#qmin# was only applied to scatterings, and not splittings. Thus, using the case of collinear splitting allows to directly study the effects of varying values for \verb#qmin# without having to consider additional effects from transverse momentum broadening via splitting.
For the parameters \verb#xeps# and \verb#xmin# parton cascades were produced for the values $10^{-4}$ and $10^{-3}$ (where always \verb#xmin#$\geq$\verb#xeps# was considered).
For the \verb#qmin# parameter, parton cascades were obtained for the values $0.1$ and $0.2$. 
From the simulated parton cascades, results were obtained for the multiplicity distributions $\frac{dN}{dk_T}$ and $\frac{dN}{dk_+}$, where $N$ is the number of produced jet particles and $k_+=xp_+$.
Results for $\frac{dN}{dk_T}$ and $\frac{dN}{dk_+}$  are shown for different values of \texttt{xeps} in Fig.~\ref{fig2}. As it can be seen the results do not exhibit large changes with regard to changes in \texttt{xeps}. As the parameter \texttt{xeps} corresponds to the infrared cut-off $\epsilon$ introduced in Eqs.~(\ref{eq:phig}) and ~(\ref{eq:phiq}) this implies that results for the parton momenta are largely independent of the selected cut-off scale.
Results for $\frac{dN}{dk_T}$ and $\frac{dN}{dk_+}$  are shown for different values of \texttt{qmin} in Fig.~\ref{fig3}. As it can be seen the results for the distribution in $k_+$ do not exhibit large changes with regard to the choice of \texttt{qmin}. On the other hand, for the distributions in $k_T$ small but noticeable differences appear (up to $15$\% at the peak of the distributions for gluons at $t_{\rm max}=1$~fm/c): The parameter $q_{\rm min}$ is a necessary -- yet artificial -- cut-off scale introduced in Eqs.~(\ref{eq:phig}) and ~(\ref{eq:phiq}) to allow for the calculation of $\Phi_g$ and $\Phi_q$. In order to allow for suitable calculations of the in-medium $k_T$ broadening it should, thus, be set to a value that is as small as possible.
Results for $\frac{dN}{dk_T}$ and $\frac{dN}{dk_+}$  are shown for different values of \texttt{xmin} in Fig.~\ref{fig4}. As it can be seen the results for the distribution in $k_+$ do not exhibit large changes with regard to the choice of \texttt{xmin}.
However, for the distributions in $k_T$ huge differences, especially at the peaks at small momentum scales occur.
In contrast to $\epsilon$ and $q_{\rm min}$, $x_{\rm min}$ is a physical scale that describes down to which scales in light-cone energy $k_+$ the phenomena of coherent medium induced radiations and scatterings apply to particles, and whether particles produced in splittings are still considered as jet, or rather as particles of a thermalized medium. 
As can be seen in Fig.~\ref{fig4}, for lower scales of $x_{\rm min}$, more soft particles are produced.
A possible choice for $x_{\rm min}$ could be given in the form of an energy scale of the order of the medium temperature $T$ at which the onset of thermalization of jet-particles is assumed. 
Another possibility would be to consider a value of $x_{\rm min}$ that corresponds to the energy-scale below which emissions of the Bethe-Heitler type rather than coherent emissions dominate.
These parametrizations of $x_{\rm min}$ depend on the other parameters for the medium and, therefore, this calibration is left to potential users that combine the \tmdice~program with their own model for the medium.

Figs.~\ref{fig1} -- \ref{fig4} represent direct qualitative results from \tmdice\;for the phenomenology of parton-cascades that undergo in the medium processes of coherent medium induced radiation and scatterings. As, in particular, the distributions of momentum components in Figs.~\ref{fig2} -- \ref{fig4} show, due to the splitting processes from highly energetic particles a large amount of soft particles is created, without accumulations at intermediate energy scales. Multiple scatterings yield a broadening in the $k_T$ distributions. Both phenomena have been discussed in more detail e.g. in~\cite{Kutak:2018dim,Blanco:2020uzy} for the case of cascades with gluons only, as well as in~\cite{Blanco:2021usa} for the cases of cascades with both, quarks and gluons. 

For a quantitative description of experimental data, inclusion of several further physical phenomena is necessary: First of all, a description for the production of jet particles in hard nuclear collisions within heavy ion collisions is necessary. 
Furthermore, before interacting with the medium via the processes of coherent medium induced radiations and scatterings, jet particles are created by the emission of bremsstrahlung, a process that can also occur within the medium as vacuum like emissions (VLEs). 
This paper leaves the necessary combinations of \tmdice~with other event generators that include the additional effects for future work. 
Instead it should be stressed that \tmdice~describes those stages of jet-evolution, where coherent medium induced radiations together with scatterings dominate. These processes yield parton multiplicity distributions that follow Eqs.~(\ref{eq:BDIM_g_mult}) and~(\ref{eq:BDIM_q_mult}) (and fragmentation functions that follow Eqs.~(\ref{eq:BDIM_g}) and~(\ref{eq:BDIM_q})), which was demonstrated numerically in this section.
\section{Conclusions and Outlook}
\label{sec4}
This paper presented the \tmdice\, program that allows to generate parton jets that are created by interactions with a constant medium from an initial jet-particle. 
The jet-medium interactions are coherent medium induced radiations off a jet particle (that reproduce the BDMPS-Z spectra) as well as scatterings off medium particles~\cite{Blaizot:2012fh,Blaizot:2013vha,Blanco:2021usa}.
Instead of a solution of the corresponding sets of evolution equations for single particle densities (such as fragmentation functions) the program provides a set of jet particles within a certain energy range and their corresponding momentum components.
This allows to use results of the program to obtain contributions from regions where coherent medium induced radiation dominate for jet-observables that depend on multiple particles.
To allow for a possible inclusion of the program in more encompassing simulations of jet-productions in heavy ion collisions, the program was written in the form of a C++ library, which makes it possible to call functions that obtain in-medium fragmentation in a region (of time and jet-particle energy, which can be set by the user) in which coherent radiation is relevant.

For the current version of the program the medium was considered to be given by a constant density of scatterers and scale of transverse momentum transfer, neglecting, thus, finite size effects of the medium. 
This kind of effects may be taken into account by future adaptations of the program.

\section*{Acknowledgement}
This work was supported by the Polish National Science Centre with the grant no.\ DEC-2017/27/B/ST2/01985. 
M.R. thanks Krzysztof Kutak, Wies\l{}aw P\l{}aczek, and Andreas van Hameren for useful remarks.
\appendix
\section{Evolution equations for multiplicities and fragmentation functions}
\label{appA}
This appendix demonstrates that the \tmdice-program allows to solve evolution equations for fragmentation functions of jets that fragment in the medium via medium induced coherent radiation and scatterings off medium particles that were previously introduced in~\cite{Blaizot:2012fh,Blaizot:2013vha,Blanco:2021usa}. To this end it will be shown that the equivalent evolution equations for the multiplicity distributions of jet-particles can be solved by the \tmdice-program.

A multiplicity distribution can be defined as
\begin{equation}
    F_i(x, \mathbf{k},t)=\frac{\partial^2 N_i (t)}{\partial x \partial \mathbf{k}}\,,
\end{equation}
where $N_i$ is the number of jet-particles of type $i$, $\mathbf{k}$ is the jet-particle momentum-component orthogonal to the jet-axis and $x$ is the ratio of jet-particle light-cone energy with the light-cone energy of an initial jet-particle.
Analogously, fragmentation functions $D_i$ of particles $i$ can be defined as
\begin{equation}
    D_i(x,\mathbf{k},t)=x\frac{\partial^2 N_i (t)}{\partial x \partial \mathbf{k}}=xF_i(x,\mathbf{k},t)\,.
    \label{eq:frag_def}
\end{equation}
In~\cite{Blaizot:2012fh,Blaizot:2013vha,Blanco:2021usa} it was found that the fragmentation functions of particles undergoing coherent medium induced radiations and scatterings off medium particles obey the following set of evolution equations
\begin{align}
\frac{\partial}{\partial t} D_g(x,\mathbf{k},t) = & \:  \alpha_s \int_0^1 dz\, \int\frac{d^2\mathbf{q}}{(2\pi)^2}\bigg[2{\cal K}_{gg}(\mathbf{Q},z,\frac{x}{z}p_+) D_g\left(\frac{x}{z},\mathbf{q},t\right) + {\cal K}_{gq}(\mathbf{Q},z,\frac{x}{z}p_+) D_{q}\left(\frac{x}{z},\mathbf{q},t\right) \nonumber\\&
-  \left({\cal K}_{gg}(\mathbf{q},z,xp_+)+{\cal K}_{qg}(\mathbf{q},z,xp_+)\right)\, D_g(x,\mathbf{k},t) \bigg] \bigg|_{\mathbf{Q}=\mathbf{k}-z\mathbf{q} }
+ \int \frac{d^2\mathbf{l}}{(2\pi)^2} \,C_g(\mathbf{l})\, D_g(x,\mathbf{k}-\mathbf{l},t)\,,\label{eq:BDIM_g}\\
\frac{\partial}{\partial t} D_{q}(x,\mathbf{k},t) = & \:  \alpha_s \int_0^1 dz\, \int\frac{d^2\mathbf{q}}{(2\pi)^2}\bigg[{\cal K}_{qq}(\mathbf{Q},z,\frac{x}{z}p_+) D_{q}\left(\frac{x}{z},\mathbf{q},t\right) +  {\cal K}_{qg}(\mathbf{Q},z,\frac{x}{z}p_+) D_g\left(\frac{x}{z},\mathbf{q},t\right) \nonumber\\&
-  {\cal K}_{qq}(\mathbf{q},z,xp_+)\, D_{q}(x,\mathbf{k},t) \bigg] \bigg|_{\mathbf{Q}=\mathbf{k}-z\mathbf{q} }
+ \int \frac{d^2\mathbf{l}}{(2\pi)^2} \,C_q(\mathbf{l})\, D_{q}(x,\mathbf{k}-\mathbf{l},t)\,,
\label{eq:BDIM_q}
\end{align}
with 
\begin{equation}
C_i(\mathbf{q}) = w_i(\mathbf{q}) - \delta(\mathbf{q}) \int d^2\mathbf{q'}\,w_i(\mathbf{q'})\,.
\label{eq:Cq}
\end{equation}
Thus, dividing Eqs.~(\ref{eq:BDIM_g}) and~(\ref{eq:BDIM_q}) with $x$ one finds that the multiplicity distributions obey the following equivalent set of evolution equations
\begin{align}
\frac{\partial}{\partial t} F_g(x,\mathbf{k},t) = & \:  \alpha_s \int_0^1 dz\, \int\frac{d^2\mathbf{q}}{(2\pi)^2}\bigg[2{\cal K}_{gg}(\mathbf{Q},z,\frac{x}{z}p_+) \frac{1}{z}F_g\left(\frac{x}{z},\mathbf{q},t\right) + {\cal K}_{gq}(\mathbf{Q},z,\frac{x}{z}p_+) \frac{1}{z}F_{q}\left(\frac{x}{z},\mathbf{q},t\right) \nonumber\\&
-  \left({\cal K}_{gg}(\mathbf{q},z,xp_+)+{\cal K}_{qg}(\mathbf{q},z,xp_+)\right)\, F_g(x,\mathbf{k},t) \bigg] \bigg|_{\mathbf{Q}=\mathbf{k}-z\mathbf{q} }
+ \int \frac{d^2\mathbf{l}}{(2\pi)^2} \,C_g(\mathbf{l})\, F_g(x,\mathbf{k}-\mathbf{l},t)\,,\label{eq:BDIM_g_mult}\\
\frac{\partial}{\partial t} F_{q}(x,\mathbf{k},t) = & \:  \alpha_s \int_0^1 dz\, \int\frac{d^2\mathbf{q}}{(2\pi)^2}\bigg[{\cal K}_{qq}(\mathbf{Q},z,\frac{x}{z}p_+) \frac{1}{z}F_{q}\left(\frac{x}{z},\mathbf{q},t\right) +  {\cal K}_{qg}(\mathbf{Q},z,\frac{x}{z}p_+) \frac{1}{z}F_g\left(\frac{x}{z},\mathbf{q},t\right) \nonumber\\&
-  {\cal K}_{qq}(\mathbf{q},z,xp_+)\, F_{q}(x,\mathbf{k},t) \bigg] \bigg|_{\mathbf{Q}=\mathbf{k}-z\mathbf{q} }
+ \int \frac{d^2\mathbf{l}}{(2\pi)^2} \,C_q(\mathbf{l})\, F_{q}(x,\mathbf{k}-\mathbf{l},t)\,.
\label{eq:BDIM_q_mult}
\end{align}
The integro-differential Eqs.~(\ref{eq:BDIM_g_mult}) and~(\ref{eq:BDIM_q_mult}) can be transformed  into the following integral equations:
\begin{align}
    F_g(x,\mathbf{k},t)=&\Delta_g(x,t-t_0)F_g(x,\mathbf{k},t_0)+\int_{t_0}^t dt' \Delta_g(x,t-t')\bigg\{ \alpha_s\int dz\int \frac{d^2 q}{(2\pi)^2} \frac{1}{z}\bigg[2\mathcal{K}_{gg}(\mathbf{Q},z,\frac{x}{z}p_+)F_g(\frac{x}{z},\mathbf{q},t')
    \nonumber\\
    &+\mathcal{K}_{gq}(\mathbf{Q},z,\frac{x}{z}p_+)F_q(\frac{x}{z},\mathbf{q},t')\bigg] \bigg|_{\mathbf{Q}=\mathbf{k}-z\mathbf{q} }+\int \frac{d^2l}{(2\pi)^2}w_g(\mathbf{l})F_g(x,\mathbf{k}-\mathbf{l},t')\bigg\}\,,
    \label{eq:bdim_int_g}
    \\
    F_q(x,\mathbf{k},t)=&\Delta_q(x,t-t_0)F_q(x,\mathbf{k},t_0)+\int_{t_0}^t dt' \Delta_q(x,t-t')\bigg\{ \alpha_s\int dz\int \frac{d^2 q}{(2\pi)^2} \frac{1}{z}\bigg[\mathcal{K}_{qq}(\mathbf{Q},z,\frac{x}{z}p_+)F_q(\frac{x}{z},\mathbf{q},t')
    \nonumber\\
    &+\mathcal{K}_{qg}(\mathbf{Q},z,\frac{x}{z}p_+)F_g(\frac{x}{z},\mathbf{q},t')\bigg] \bigg|_{\mathbf{Q}=\mathbf{k}-z\mathbf{q} }+\int \frac{d^2l}{(2\pi)^2}w_q(\mathbf{l})F_q(x,\mathbf{k}-\mathbf{l},t')\bigg\}\,,
    \label{eq:bdim_int_q}
\end{align}
Finally these equations can be written in the following form
\begin{align}
    F_g(x,\mathbf{k},t)=&\Delta_g(x,t-t_0)F_g(x,\mathbf{k},t_0)+\int_{t_0}^t dt' \Delta_g(x,t-t')\bigg\{ \alpha_s\int dz \int dy \int \frac{d^2 \mathbf{q}}{(2\pi)^2}\int d^2\mathbf{Q} \delta(x-zy)        \nonumber\\
    &\delta^2(\mathbf{k}-\mathbf{Q}-z\mathbf{q})
\bigg[2\frac{\mathcal{K}_{gg}(\mathbf{Q},z,yp_+)}{\rho_{gg}(y)}\frac{\rho_{gg}(y)}{\rho_{gg}(y)+\rho_{qg}(y)}\frac{\rho_{gg}(y)+\rho_{qg}(y)}{\phi_g(y)}\phi_g(y)F_g(y,\mathbf{q},t')
    \nonumber\\
+&\frac{\mathcal{K}_{gq}(\mathbf{Q},z,yp_+)}{\rho_{qq}(y)}\frac{\rho_{qq}(y)}{\phi_q(y)}\phi_q(y)F_q(y,\mathbf{q},t')\bigg]
    \nonumber\\
    &+\int dy \int d^2\mathbf{Q}  \frac{d^2\mathbf{q}}{(2\pi)^2} \delta(y-x) \delta^2(\mathbf{k}-\mathbf{q}-\mathbf{Q})\frac{w_g(\mathbf{q})}{W_g}  
    \left( \frac{W_g}{\phi_g(y)}\right)\phi_g(y)
    F_g(y,\mathbf{Q},t')\bigg\}\,,
    \label{eq:bdim_int_g3}
    \\
    F_q(x,\mathbf{k},t)=&\Delta_q(x,t-t_0)F_q(x,\mathbf{k},t_0)+\int_{t_0}^t dt' \Delta_q(x,t-t')\bigg\{ \alpha_s\int dy \int dz\int d^2\mathbf{Q}\int \frac{d^2 \mathbf{q}}{(2\pi)^2} \delta(x-zy)        \nonumber\\
    &\delta^2(\mathbf{k}-\mathbf{Q}-z\mathbf{q})
\bigg[\frac{\mathcal{K}_{qq}(\mathbf{Q},z,yp_+)}{\rho_{qq}(y)}\frac{\rho_{qq}(y)}{\phi_q(y)}\phi_q(y)F_q(y,\mathbf{q},t')
    \nonumber\\
+&\frac{\mathcal{K}_{qg}(\mathbf{Q},z,yp_+)}{\rho_{qg}(y)}\frac{\rho_{qg}(y)}{\rho_{qg}(y)+\rho_{gg}(y)}\frac{\rho_{qg}(y)+\rho_{gg}(y)}{\phi_g(y)}\phi_g(y)F_g(y,\mathbf{q},t')\bigg]
    \nonumber\\
    &+\int dy \int d^2\mathbf{Q} \int  \frac{d^2\mathbf{q}}{(2\pi)^2} \delta(y-x) \delta^2(\mathbf{k}-\mathbf{q}-\mathbf{Q}) \frac{w_q(\mathbf{q})}{W_q}  
    \frac{W_q}{\phi_q(y)}\phi_q(y)F_q(y,\mathbf{Q},t')\bigg\}\,,
    \label{eq:bdim_int_q3}
\end{align}
where $\rho_{gq}(y)=\rho_{qq}(y)$ was used.
The above set of evolution equations has an iterative solution, which can be obtained by substituting the equations  into the multiplicity distributions  at the right sides that are convoluted with the scattering and splitting kernels and repeating this procedure multiple times. 
Thus, one obtains
\begin{align}
    F_g(x,\mathbf{k},t)=&\Delta_g(x,t-t_0)F_g(x,\mathbf{k},t_0)+\int_{t_0}^t dt' \Delta_g(x,t-t')\bigg\{ \alpha_s\int dz \int dy \int \frac{d^2 \mathbf{q}}{(2\pi)^2}\int d^2\mathbf{Q} \delta(x-zy)        \nonumber\\
    &\delta^2(\mathbf{k}-\mathbf{Q}-z\mathbf{q})
\bigg[2\frac{\mathcal{K}_{gg}(\mathbf{Q},z,yp_+)}{\rho_{gg}(y)}\frac{\rho_{gg}(y)}{\rho_{gg}(y)+\rho_{qg}(y)}\frac{\rho_{gg}(y)+\rho_{qg}(y)}{\phi_g(y)}\phi_g(y)\Delta_g(y,t'-t_0)F_g(y,\mathbf{q},t_0)
    \nonumber\\
+&\frac{\mathcal{K}_{gq}(\mathbf{Q},z,yp_+)}{\rho_{qq}(y)}\frac{\rho_{qq}(y)}{\phi_q(y)}\phi_q(y)\Delta_q(y,t'-t_0)F_q(y,\mathbf{q},t_0)\bigg]
    \nonumber\\
    &+\int dy \int d^2\mathbf{Q}  \frac{d^2\mathbf{q}}{(2\pi)^2} \delta(y-x) \delta^2(\mathbf{k}-\mathbf{q}-\mathbf{Q})\frac{w_g(\mathbf{q})}{W_g}  
    \left( \frac{W_g}{\phi_g(y)}\right)\phi_g(y)
    \Delta_g(y,t'-t_0)F_g(y,\mathbf{Q},t_0)\bigg\}
    \nonumber\\&+\mathcal{O}(\alpha_s^2)\,,
    \label{eq:bdim_int_g4}
    \\
    F_q(x,\mathbf{k},t)=&\Delta_q(x,t-t_0)F_q(x,\mathbf{k},t_0)+\int_{t_0}^t dt' \Delta_q(x,t-t')\bigg\{ \alpha_s\int dy \int dz\int d^2\mathbf{Q}\int \frac{d^2 \mathbf{q}}{(2\pi)^2} \delta(x-zy)
        \nonumber\\
        &\delta^2(\mathbf{k}-\mathbf{Q}-z\mathbf{q})
    \bigg[\frac{\mathcal{K}_{qq}(\mathbf{Q},z,yp_+)}{\rho_{qq}(y)}\frac{\rho_{qq}(y)}{\phi_q(y)}\phi_q(y)\Delta_q(y,t'-t_0)F_q(y,\mathbf{q},t_0)
    \nonumber\\
+&\frac{\mathcal{K}_{qg}(\mathbf{Q},z,yp_+)}{\rho_{qg}(y)}\frac{\rho_{qg}(y)}{\rho_{qg}(y)+\rho_{gg}(y)}\frac{\rho_{qg}(y)+\rho_{gg}(y)}{\phi_g(y)}\phi_g(y)\Delta_g(y,t'-t_0)F_g(y,\mathbf{q},t_0)\bigg]
    \nonumber\\
    &+\int dy \int d^2\mathbf{Q} \int  \frac{d^2\mathbf{q}}{(2\pi)^2} \delta(y-x) \delta^2(\mathbf{k}-\mathbf{q}-\mathbf{Q}) \frac{w_q(\mathbf{q})}{W_q}  
    \frac{W_q}{\phi_q(y)}\phi_q(y)\Delta_q(y,t'-t_0)F_q(y,\mathbf{Q},t_0)\bigg\}
    \nonumber\\&
    +\mathcal{O}(\alpha_s^2)\,,
    \label{eq:bdim_int_q4}
\end{align}
where only the contributions from one splitting or scattering were written explicitly and contributions from more than one process of jet-medium interaction was denoted by $\mathcal{O}(\alpha_s^2)$.
From the above formulation of the system of integral evolution equations a possible solution via Monte-Carlo algorithms becomes apparent: 
Initial Monte-Carlo samples for the light cone energy fractions and the transverse momenta are selected from the multiplicity distributions $F_q$ and $F_g$ at time $t_0$. 
Possible changes to the Monte-Carlo samples during the in-medium fragmentation 
follow the same probabilities and probability densities that were given earlier on in Eqs.~(\ref{eq:interactiontime})-(\ref{eq:scatq}).
Thus, it follows that the \tmdice-algorithm described in this paper provides a Monte-Carlo solution to Eqs.~(\ref{eq:BDIM_g_mult}) and~(\ref{eq:BDIM_q_mult}) (provided the initial particles are properly from an initial condition at times $t_0$).
The corresponding fragmentation functions can be obtained via Eq.~(\ref{eq:frag_def}) and are solutions to Eqs.~(\ref{eq:BDIM_g}) and~(\ref{eq:BDIM_q}).

\bibliographystyle{elsarticle-num}
%\bibliography{refs}{}

\end{document}